\documentclass[12pt]{article}
\pdfoutput=1
\usepackage[nosort]{cite}

\usepackage{xcolor}

\usepackage{epsfig}
\usepackage{amsfonts}
\usepackage{amscd}
\usepackage{latexsym}
\usepackage{amsmath,amssymb}
\usepackage{amsthm}
\usepackage{verbatim}
\usepackage{setspace}
\usepackage{color}
\usepackage{fancyhdr}
\usepackage{tikz}
\usepackage{tikz-cd}
\usepackage{slashed}
\usepackage{soul}
\usepackage{multirow}

\usetikzlibrary{calc}
\usetikzlibrary{topaths}
\usetikzlibrary{decorations}
\usetikzlibrary{decorations.pathmorphing}

\usetikzlibrary{zx-calculus}

\usepackage{draft}
\usepackage{hyperref}
\usepackage{graphicx,color,subfig}
\usepackage{cite}
\usepackage{mciteplus}
\usepackage{skak}
\usepackage{centernot}

\usepackage{enumitem}
\usepackage{crossreftools}

\numberwithin{equation}{section}

\renewcommand{\b}[1]{\boldsymbol{#1}}

\newcommand{\im}{\operatorname{im}}

\newcommand{\Plus}{\mathord{\tikz[scale=1.5,baseline=1pt]\draw[line width=0.25ex, x=1ex, y=1ex] (0.5,0) -- (0.5,1)(0,0.5) -- (1,0.5);}}
\newcommand{\Zero}{{\mathsf 0}}
\newcommand{\One}{{\mathsf 1}}
\newcommand{\mmm}{{\mathsf m}}

\def\({\left(}
\def\){\right)}

\def\>{\rangle}
\def\<{\langle}

\begin{document}

\begin{titlepage}
	

\hfill \\
\hfill \\
\vskip 1cm

\title{Duality-preserving deformation of 3+1d lattice $\mathbb Z_2$ gauge theory with exact gapped ground states}

\author{Pranay Gorantla$^{1}$ and Tzu-Chen Huang$^{1}$}

\address{${}^{1}$Kadanoff Center for Theoretical Physics \& Enrico Fermi Institute, University of Chicago}


\vspace{2.0cm}

\begin{abstract}\noindent
We propose and analyze a deformation of the 3+1d lattice $\mathbb Z_2$ gauge theory that preserves the non-invertible Wegner duality symmetry at the self-dual point. We identify a frustration-free point along this deformation where there are nine exactly degenerate ground states (on a periodic cubic lattice) even at finite volume. One of these ground states is a trivial product state and the rest are the topologically-ordered ground states of the 3+1d toric code. We also prove that the frustration-free point is gapped in the thermodynamic limit. Our model, therefore, realizes a gapped phase with spontaneously broken Wegner duality symmetry. Furthermore, by imposing the Gauss law constraints energetically, all the above features can be realized on a tensor product Hilbert space. Finally, we discuss a generalization of this deformation to the 3+1d lattice $\mathbb Z_N$ gauge theory and conjecture the possible phase diagram.
\end{abstract}

\vfill
	
\end{titlepage}

\eject

\tableofcontents

\section{Introduction}\label{sec:intro}
Recently, there has been a growing interest in realizing generalized global symmetries, including non-invertible symmetries, on the lattice---see \cite{Inamura:2021szw,Tan:2022vaz,Eck:2023gic,Mitra:2023xdo,Sinha:2023hum,Fechisin:2023dkj,Yan:2024eqz,Okada:2024qmk,Seifnashri:2024dsd,Bhardwaj:2024wlr,Chatterjee:2024ych,Bhardwaj:2024kvy,Khan:2024lyf,Jia:2024bng,Lu:2024ytl,Li:2024fhy,Ando:2024hun,ODea:2024tkt,Pace:2024tgk} for 1+1d examples and \cite{Delcamp:2023kew,Inamura:2023qzl,Moradi:2023dan,Cao:2023doz, Cao:2023rrb,ParayilMana:2024txy,Spieler:2024fby,Choi:2024rjm,Hsin:2024aqb,Cao:2024qjj} for 2+1d examples. The presence of such symmetries yields nontrivial information about the possible low energy phases of lattice models via LSM-type constraints \cite{Levin:2019ifu,Seiberg:2024gek}. On the flip side, such lattice models offer a fruitful playground for exploring aspects of generalized symmetries not seen in the continuum (e.g. mixing with spatial symmetries such as lattice translations, etc. \cite{Seiberg:2023cdc,Seiberg:2024gek,Gorantla:2024ocs}).

Arguably, the simplest lattice model with a non-invertible symmetry is the 1+1d transverse-field Ising model. At the critical point, in addition to the $\mathbb Z_2$ symmetry, it has a non-invertible symmetry associated with the Kramers-Wannier duality \cite{Grimm:1992ni,Oshikawa:1996dj,Ho:2014vla,Hauru:2015abi,Aasen:2016dop}. This non-invertible symmetry is not compatible with a trivially gapped phase, which is interpreted as an LSM-type constraint \cite{Levin:2019ifu,Seiberg:2024gek}. Instead, it is compatible with (i) a gapless phase or (ii) a gapped phase with three ground states where the Kramers-Wannier duality symmetry is spontaneously broken. The critical point of the transverse-field Ising model realizes option (i).

One might wonder if there is a lattice model that realizes option (ii). In \cite{OBrien:2017wmx}, O'Brien and Fendley proposed a Kramers-Wannier duality-preserving deformation of the critical Ising model. (See also \cite{Sannomiya:2017foz} for a similar deformation.) Their numerical analysis suggests that, along this deformation, there is a transition from a gapless phase realizing option (i) to a gapped phase realizing option (ii). Interestingly, there is a ``frustration-free''\footnote{A frustration-free Hamiltonian is a local Hamiltonian such that (i) each local term is positive semi-definite and (ii) there are zero energy ground states. The first property implies that all energies are nonnegative, and the second property implies that the zero energy states (i.e., the ground states) must be annihilated by each and every local term of the Hamiltonian.} point within the gapped phase of the deformation which hosts three exactly degenerate ground states even at finite volume.

A natural generalization of Kramers-Wannier duality is the Wegner duality in the 3+1d lattice $\mathbb Z_2$ gauge theory \cite{Wegner:1971app}. The self-dual point of the lattice $\mathbb Z_2$ gauge theory separates the trivial and topologically-ordered gapped phases. At this point, in addition to the $\mathbb Z_2$ 1-form symmetry, there is a non-invertible symmetry associated with the Wegner duality. The operator that generates this non-invertible symmetry on the lattice, as well as its operator algebra, was analyzed recently in \cite{Gorantla:2024ocs}. (See also \cite{Koide:2021zxj,Choi:2021kmx,Kaidi:2021xfk} for related discussions in Euclidean lattice gauge theory and in continuum field theory.) Like the Kramers-Wannier duality symmetry, the Wegner duality symmetry is not compatible with a trivially gapped phase; instead, it is compatible with (i) a gapless phase or (ii) a gapped phase where the Wegner duality symmetry is spontaneously broken \cite{Gorantla:2024ocs}.\footnote{It is not yet known if there is a gapped phase with unbroken Wegner duality symmetry.}

In this work, we propose a deformation (parametrized by $\lambda \ge 0$) of the 3+1d lattice $\mathbb Z_2$ gauge theory that preserves the non-invertible Wegner duality symmetry at the self-dual point. This deformation was advertised in \cite{Gorantla:2024ocs}, and is analogous to O'Brien and Fendley's deformation of the critical Ising model \cite{OBrien:2017wmx}. We mainly focus on a particular point along this deformation ($\lambda = 1$) where the model is ``frustration-free'' and realizes option (ii) of the last paragraph. Some prominent features of this frustration-free point are listed below:
\begin{enumerate}
\item there are nine ground states on a periodic cubic lattice, all of which can be computed explicitly: one is a trivial product state and the rest are the eight topologically-ordered ground states of the 3+1d toric code,
\item the ground states are exactly degenerate even at finite volume, and
\item there is a nonzero gap in the thermodynamic limit.
\end{enumerate}
While the second statement follows trivially from frustration-freeness of our model, the first and third statements are much harder to prove. The bulk of this paper is devoted to proving them.

We use a combinatorial argument in proving the nine-fold degeneracy. First, we write an arbitrary ground state as a superposition of sign configurations (i.e., eigenstates of Pauli $X$ operators) on the links of the cubic lattice. Frustration-freeness implies that two sign configurations in the superposition have equal weights if they are related by a sequence of local flips, which are reminiscent of moves that relate homologous curves. Using this connection to homology, we show that two configurations are related by a sequence of local flips if and only if they carry the same $\mathbb Z_2$ 1-form symmetry charges, which then implies the desired nine-fold degeneracy.

Coming to the gap, in general, it is impossible to prove the gap of an arbitrary local Hamiltonian \cite{Cubitt:2015xsa,Cubitt:2015lta}. But for frustration-free Hamiltonians, there are a couple of methods that are commonly used: the Knabe method \cite{Knabe:1988,Gosset:2016} and the martingale method \cite{Fannes:1992,Nachtergaele:1996,Kastoryano_2018}. The former requires numerical computation, which is impractical in higher dimensions, whereas the latter requires the explicit knowledge of ground states. Given that we know the ground states explicitly, we use the latter to prove the gap.

To appreciate the nontriviality of the above statements, let us compare our model with a well-known class of lattice models: commuting projector Hamiltonians.\footnote{A local Hamiltonian is commuting projector if all the local terms of the Hamiltonian commute with each other. Note that a commuting-projector Hamiltonian is not necessarily frustration-free. For example, the antiferromagnetic Ising model (without the transverse-field term) on a triangular lattice is commuting-projector but not frustration-free.} When a frustration-free Hamiltonian is also commuting-projector, then it is typically exactly solvable and the gap is trivially independent of the system size. Many commonly encountered Hamiltonians with topological order, such as the toric code, are of this type. Our model, however, is not commuting-projector, so it is not exactly solvable.\footnote{While there are exactly solvable models that are not commuting projector, such as integrable models in 1+1d, they are not common in higher dimensions.} Nevertheless, thanks to frustration-freeness, we are able to compute the ground states explicitly and prove that there is a nonzero gap between the ground states and the first excited state in the thermodynamic limit.

Let us also compare our model ($\lambda = 1$) with the self-dual point of the lattice $\mathbb Z_2$ gauge theory ($\lambda = 0$). Numerical studies suggest that, at the self-dual point, there is a first-order transition between trivial (confining) and topologically-ordered (Higgs) phases \cite{PhysRevLett.42.1390,PhysRevD.20.1915}, so it realizes option (ii) as well. However, the self-dual point is not frustration-free. So the ground states are not known explicitly, they are degenerate only in the infinite volume limit, and the gap suggested by numerics is hard to prove analytically. This is in stark contrast to the features of our model listed above.

Finally, we emphasize that all the above features can be realized on a tensor product Hilbert space by imposing the Gauss law constraints energetically rather than exactly.

The rest of this paper is organized as follows. In Section \ref{sec:3d-review}, we briefly review the symmetries and the phase diagram of the 3+1d lattice $\mathbb Z_2$ gauge theory. In Section \ref{sec:3d-deform}, we introduce the deformation that preserves the Wegner duality symmetry at the self-dual point, and specialize to a particular frustration-free point along this deformation. We then prove the exact nine-fold degeneracy at this point and analytically prove the gap in the thermodynamic limit. Finally, in Section \ref{sec:discuss}, we summarize our results and discuss a generalization of our deformation to the lattice $\mathbb Z_N$ gauge theory.\footnote{We focus on $\mathbb Z_2$ in the main text for simplicity of presentation. All our proofs and results generalize straightforwardly to $\mathbb Z_N$.} We also make some brief comments on the possible schematic phase diagram of the deformed lattice $\mathbb Z_N$ gauge theory. There are also three appendices. Appendices \ref{app:deformed-nogausslaw} and \ref{app:up-bound-delta} contain some technical details used in the proofs in Section \ref{sec:3d-deform}. In Appendix \ref{app:deformed-Ising}, we review the Kramers-Wannier duality-preserving deformation of the 1+1d critical Ising model \cite{OBrien:2017wmx}, give an alternative proof of the three-fold degeneracy of the frustration-free point, and provide an analytic proof of gap in the thermodynamic limit.

\section{Brief review of 3+1d lattice $\mathbb Z_2$ gauge theory}\label{sec:3d-review}
Consider an $L_x \times L_y \times L_z$ periodic cubic lattice with $V:=L_x L_y L_z$ sites. We use $s$, $\ell$, $p$, and $c$ to denote the sites, links, plaquettes, and cubes of the lattice. We also use $\b x = (x,y,z) \in (\mathbb Z/L_x \mathbb Z) \times (\mathbb Z/L_y \mathbb Z) \times (\mathbb Z/L_z \mathbb Z)$ to label the sites, and appropriately shifted half-integral coordinates to represent links, plaquettes, and cubes.

Each link hosts a qubit, so the local Hilbert space is $\mathcal H_\ell \cong \mathbb C^2$. Let $\mathcal H$ be the tensor product Hilbert space $\bigotimes_\ell \mathcal H_\ell$. Define the constrained Hilbert space $\widetilde{\mathcal H}$ as the subspace of $\mathcal H$ satisfying the Gauss law constraint,
\ie\label{gausslawconstraint}
G_s := \prod_{\ell \ni s} X_\ell = 1~,
\fe
at each site $s$. Equivalently, $\widetilde{\mathcal H}$ is the image of the orthogonal projection $\prod_s \left( \frac{1+G_s}{2} \right)$.

\begin{figure}
\centering
\hfill \raisebox{-0.5\height}{\includegraphics[scale=0.25]{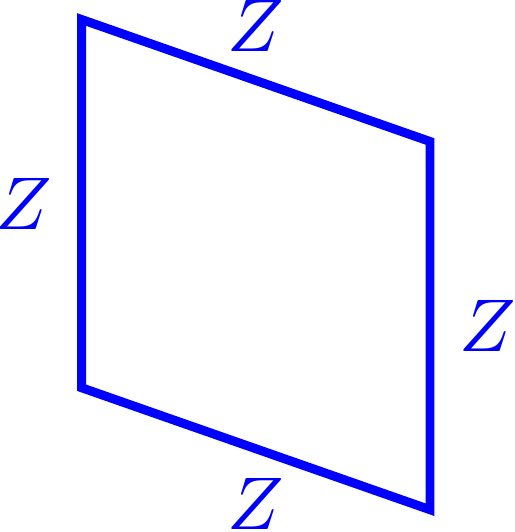}} \hfill \raisebox{-0.5\height}{\includegraphics[scale=0.25]{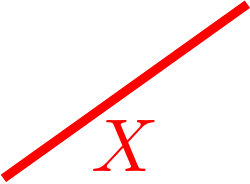}} \hfill $G_s = $~\raisebox{-0.5\height}{\includegraphics[scale=0.25]{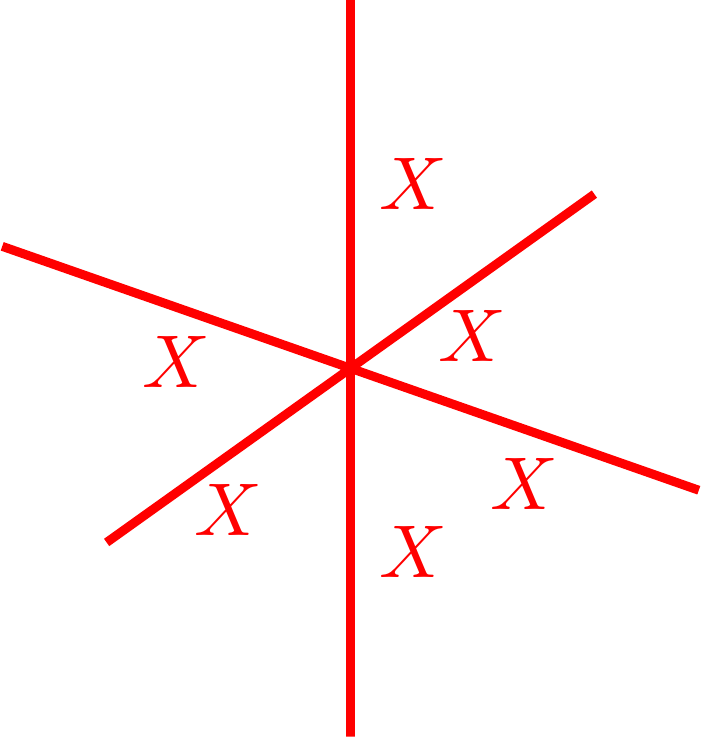}} \hfill ~
\\~\\
\hfill (a)~ \hfill (b)~~~~~~ \hfill (c) \hfill ~~~~~~~~~
\caption{The local terms of the Hamiltonian \eqref{tH} of the lattice $\mathbb Z_2$ gauge theory: (a) the magnetic flux term and (b) the electric field term. (c) The Gauss law operator $G_s$.\label{fig:Hterms}}
\end{figure}

The Hamiltonian of the 3+1d lattice $\mathbb Z_2$ gauge theory is
\ie\label{tH}
\widetilde H = -J \sum_p \prod_{\ell\in p} Z_\ell - h \sum_\ell X_\ell~.
\fe
The first term is known as the magnetic flux term, whereas the second term is known as the electric field term. (These local terms and the Gauss law operator are illustrated in Figure \ref{fig:Hterms}.) In this presentation, we impose the Gauss law exactly, so the total Hilbert space is $\widetilde{\mathcal H}$.

Alternatively, one can impose the Gauss law energetically by modifying the Hamiltonian to
\ie\label{H}
H = -J \sum_p \prod_{\ell\in p} Z_\ell - h \sum_\ell X_\ell - g \sum_s G_s~.
\fe
In this case, the total Hilbert space is $\mathcal H$. Since $G_s$ commutes with all the terms in the Hamiltonian, any state that violates the Guass law constraints incurs an energy cost proportional to $g$. Therefore, for $g>0$, the low-energy physics of both models is the same; in particular, the ground states of $H$ are the same as the ground states of $\widetilde H$.

\subsection{Symmetries}
Evidently, both $\widetilde H$ and $H$ are invariant under lattice translations and rotations. They also have a $\mathbb Z_2$ 1-form symmetry (denoted as $\mathbb Z_2^{(1)}$) generated by the operators
\ie
\eta(\widehat \Sigma) = \prod_{\ell \in \widehat \Sigma} X_\ell~,
\fe
where $\widehat \Sigma$ is a dual surface on the dual lattice, and $\ell \in \widehat \Sigma$ means $\ell$ pierces the dual surface $\widehat \Sigma$. The 1-form symmetry of $\widetilde H$ is ``topological'' in the sense that
\ie
\widehat \Sigma_1 \sim \widehat \Sigma_2 \implies \eta(\widehat \Sigma_1) = \eta(\widehat \Sigma_2)~,
\fe
i.e., the 1-form symmetry operators associated with homologous dual surfaces act identically on any state in $\widetilde{\mathcal H}$. This follows from the Gauss law constraint obeyed by all states in $\widetilde{\mathcal H}$. In contrast, the 1-form symmetry of $H$ is not topological due to the presence of Gauss law violating states in $\mathcal H$.

When $J=h$, in addition to the above invertible symmetries, there is a non-invertible Wegner duality symmetry generated by the operator $\mathsf D$ that exchanges the electric and magnetic terms of the Hamiltonian:\footnote{It is non-invertible in the following sense. $\mathsf D$ maps the symmetry operator $\eta(\widehat \Sigma)$ to the identity operator, i.e, $\mathsf D \eta(\widehat \Sigma) = \mathsf D$. If $\mathsf D$ were invertible, this action would imply $\eta(\widehat \Sigma) = 1$, which is a contradiction.}
\ie
\mathsf D X_\ell = \left(\prod_{\ell' \in \mathfrak t(\ell)} Z_{\ell'}\right) \mathsf D~,\qquad \mathsf D \left(\prod_{\ell\in p} Z_\ell\right) = X_{\mathfrak t(p)} \mathsf D~.
\fe
Here, $\mathfrak t$ is a \emph{half-translation} map given by
\ie
\mathfrak t(x,y,z) = (x+\tfrac12, y+\tfrac12, z+\tfrac12)~.
\fe
In particular, $\mathfrak t$ maps links (resp. plaquettes) to plaquettes (resp. links) that are $(\tfrac12,\tfrac12,\tfrac12)$ away. The expression of $\mathsf D$ and its operator algebra with other symmetry operators were analysed recently in \cite{Gorantla:2024ocs} using a tensor network formalism known as ZX-calculus \cite{Coecke:2008lcg,Duncan:2009ocf,vandeWetering:2020giq}. 

\subsection{Ground states and superselection sectors}\label{sec:3d-groundstates}
Numerics suggests that the 3+1d lattice $\mathbb Z_2$ gauge theory has two gapped phases \cite{PhysRevLett.42.1390,PhysRevD.20.1915}: a topologically ordered (Higgs) phase for $J>h$ and a trivial (confining) phase for $J<h$. The transition between the two phases is first-order and occurs precisely at $J=h$ due to the Wegner duality.

The confining phase has a unique ground state---for instance, at $J=0$, the ground state is given by the product state
\ie\label{3d-prodstate}
|{+}\cdots{+}\>~.
\fe
On the other hand, on a periodic cubic lattice, there are $2^3 = 8$ ground states in the Higgs phase---in particular, at $h=0$, the eight ground states are given by
\ie\label{3d-toriccodestates}
|\xi\> := 2^{(V-1)/2} \prod_{i<j} \eta_{ij}^{\xi_{ij}} \prod_s \left( \frac{1+G_s}{2} \right) |0\cdots0\>~, 
\fe
where $\xi := (\xi_{xy},\xi_{yz},\xi_{zx}) \in \{0,1\}^3$ and $\eta_{ij}$ is the 1-form symmetry operator on the non-contractible $ij$-plane.\footnote{Due to the presence of the projection $\prod_{s} \left(\frac{1+G_s}{2}\right)$, the choice of the $ij$-plane is irrelevant.} Note that these are precisely the ground states of the 3+1d toric code.\footnote{The Hamiltonian of the 3+1d toric code is obtained by setting $h=0$ in \eqref{H}. As explained around that equation, for $g>0$, the ground states of $H$ and $\widetilde H$ are the same.} While the 1-form symmetry operators permute the states within a phase, the Wegner duality operator exchanges the two phases, i.e.,
\ie\label{3d-actiononstates}
&\prod_{i<j}\eta_{ij}^{\xi_{ij}'} |\xi\> = |\xi+\xi' \>~,\qquad \eta_{ij} |{+}\cdots{+}\> = |{+}\cdots{+}\>~,
\\
&\mathsf D |\xi\> = \frac1{\sqrt2} |{+}\cdots{+}\>~,\qquad \mathsf D |{+}\cdots{+}\> = \frac1{\sqrt2} \sum_{\xi \in \{0,1\}^3}|\xi\>~,
\fe
where the sum $\xi+\xi'$ is modulo 2.

If these nine states were ground states of a gapped Hamiltonian, then that model would realise a spontaneously broken Wegner duality symmetry as implied by \eqref{3d-actiononstates}. We construct precisely such a Hamiltonian in Section \ref{sec:3d-deform}.

But before that, let us quickly review another basis for the above states. The basis $|\xi\>$ corresponds to the ``superselection sectors'' in the topologically-ordered phase. One can also construct a basis in which the symmetry operators are diagonal. For this, we define
\ie\label{zeta-states}
|\zeta\> &:= \frac1{2\sqrt2} \sum_{\xi \in \{0,1\}^3} (-1)^{\sum_{i<j} \zeta_{ij} \xi_{ij}} |\xi\>
\\
&=2\cdot 2^{V/2} \prod_{i<j} \left( \frac{1+(-1)^{\zeta_{ij}}\eta_{ij}}{2} \right) \prod_s \left( \frac{1+G_s}{2} \right) |0\cdots0\>~,
\fe
where $\zeta := (\zeta_{xy},\zeta_{yz},\zeta_{zx}) \in \{0,1\}^3$. Then, the eigenstates of the symmetry operators and their respective eigenvalues are
\ie
&\frac1{\sqrt2} |{+}\cdots{+}\> \pm \frac1{\sqrt2} |\zeta=0\>~,\quad &&\mathsf D = \pm 2~,\quad \eta_{ij} = 1~,
\\
&|\zeta\ne0\>~,\quad &&\mathsf D = 0~,\quad \eta_{ij} = (-1)^{\zeta_{ij}}~.
\fe
Another useful representation of the states $|\zeta\>$ is\footnote{The equality of the two representations \eqref{zeta-states} and \eqref{zeta-loopgas} of $|\zeta\>$ can be shown, for example, using ZX-calculus \cite{Gorantla:2024ocs}.}
\ie\label{zeta-loopgas}
|\zeta\> = 2^{V-1} \prod_{i,j,k\atop\text{cyclic}} W_k^{\zeta_{ij}} \prod_p \left( \frac{1+\prod_{\ell \in p} Z_\ell}{2} \right) |{+}\cdots{+}\>~,
\fe
where $W_k := \prod_{\ell \in \gamma_k} Z_\ell$ is the Wilson line operator along the non-contractible cycle $\gamma_k$ in the $k$-direction.\footnote{The choice of $\gamma_k$ is irrelevant because of the presence of the projection $\prod_p \left( \frac{1+\prod_{\ell \in p} Z_\ell}{2} \right)$.}

\section{Wegner duality-preserving deformation}\label{sec:3d-deform}

In this section, we introduce and analyze the deformation of the 3+1d lattice $\mathbb Z_2$ gauge theory that preserves both the $\mathbb Z_2$ 1-form symmetry and the Wegner duality symmetry at the self-dual point.

\begin{figure}
\centering
\raisebox{-0.5\height}{\includegraphics[scale=0.25]{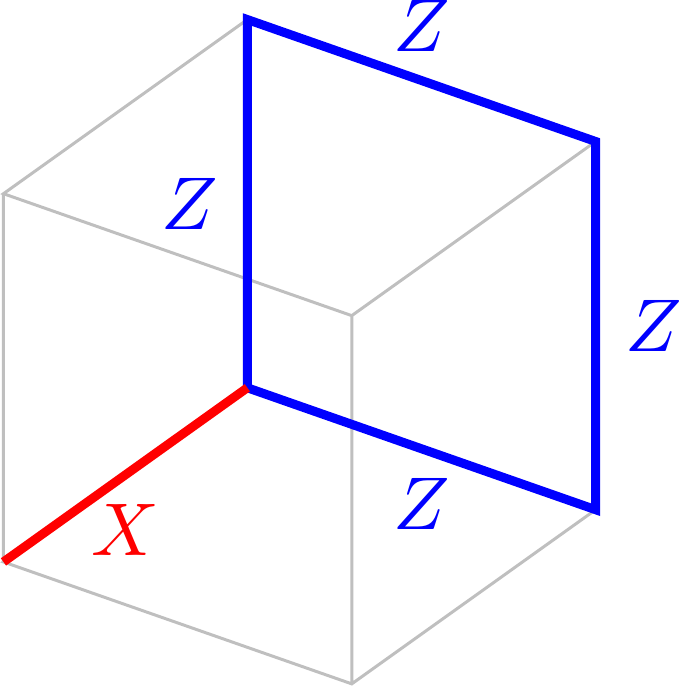}}
\caption{The deformation term in $\widetilde H_\lambda$ \eqref{deformtH}.\label{fig:deformHterm}}
\end{figure}

Consider the deformed Hamiltonian,
\ie\label{deformtH}
\widetilde H_\lambda = -J \left( \sum_p \prod_{\ell\in p}Z_\ell + \frac{h}{J}\sum_\ell X_\ell 
- \frac{\lambda}{8} \sum_{\ell,p: \ell \perp p} X_\ell \prod_{\ell'\in p} Z_{\ell'} \right)~,
\fe
where $\ell\perp p$ (or $p\perp \ell$) means the link $\ell$ is orthogonal to the plaquette $p$ and they meet at a site. See Figure \ref{fig:deformHterm} for an illustration of this deformation term. Clearly, it preserves the lattice translations and rotations, and the $\mathbb Z_2$ 1-form symmetry. When $J=h$, it also preserves the Wegner duality symmetry $\mathsf D$ because $\ell\perp p \iff \mathfrak t(\ell)\perp \mathfrak t(p)$.

This deformation is analogous to the Kramers-Wannier duality-preserving deformation of the 1+1d critical Ising model in \cite{OBrien:2017wmx}. (The latter is reviewed in Appendix \ref{app:deformed-Ising}, along with a different proof of three-fold degeneracy, as well as a proof of gap in the thermodynamic limit.)

\subsection{Exact ground states}
Let us focus on the point $\lambda = 1$ along the self-dual line $J=h$. First, it is convenient to write the Hamiltonian at this point as
\ie\label{3d-deformedtH'}
\widetilde H' = \frac{J}{8} \sum_{\ell,p:\ell\perp p} (1-X_\ell)\left( 1-\prod_{\ell'\in p}Z_{\ell'} \right) = \frac{J}{2} \sum_{\ell,p:\ell\perp p} P_\ell Q_p~,
\fe
where $P_\ell := \frac12 (1-X_\ell)$ and $Q_p := \frac12 ( 1-\prod_{\ell'\in p}Z_{\ell'} )$ are orthogonal projection operators that commute whenever $\ell\notin p$. It differs from $\widetilde H_{\lambda = 1}$ only by a scalar multiple of the identity operator. Observe that each term in $\widetilde H'$ is an orthogonal projection operator (up to the factor of $J/2$). This means each term is positive semi-definite, so $\widetilde H'$ is also positive semi-definite, and hence its energies are all nonnegative.

Moreover,
\ie\label{groundstates}
\widetilde H'|{+}\cdots{+}\> = 0~,\qquad \widetilde H'|\xi\> = 0~,
\fe
where the eight states $|\xi\>$ are defined in \eqref{3d-toriccodestates}. Therefore, there are at least $1+2^3 = 9$ ground states of $\widetilde H'$ with zero energy. In particular, this means that $\widetilde H'$ is frustration-free, i.e., any other ground state should also have zero energy and must be annihilated by every term in the Hamiltonian.

As we discussed in Section \ref{sec:3d-groundstates}, $|\xi\>$ are the ground states of the lattice $\mathbb Z_2$ gauge theory \eqref{tH} when $h=0$ (or equivalently, they  are the ground states of the 3+1d toric code) and $|{+}\cdots{+}\>$ is the ground state when $J=0$. What we have here is a frustation-free model $\widetilde H'$ that preserves the $\mathbb Z_2$ 1-form symmetry and the Wegner duality symmetry, and has these nine states as its exact ground states even at finite volume.

In the rest of this section, we show that the deformed model \eqref{3d-deformedtH'} has no other ground states, and it is gapped in the thermodynamic limit.

\subsection{Proof of nine-fold degeneracy}\label{sec:3d-proofofgsd}
We now prove that \eqref{groundstates} are the only ground states of $\widetilde H'$. Our proof is conceptually very similar to the one in Appendix \ref{sec:anotherproofGSD1d} for an analogous deformation of the 1+1d critical Ising model.

Let $|\psi\>$ be a ground state, i.e., $\widetilde H'|\psi\> = 0$. Since each term in $\widetilde H'$ is positive semi-definite, we have
\ie\label{eachtermzero3d}
P_\ell Q_p |\psi\> = 0~,
\fe
for every $\ell,p$ such that $\ell\perp p$. Let us decompose $|\psi\>$ in the eigenbasis of $X_\ell$'s:
\ie\label{arbitrarystate'}
|\psi\> = \sideset{}{'}\sum_{\sigma\in\{+,-\}^{3V}} \psi_\sigma|\sigma\>~,
\fe
where $\sigma$ denotes a sign configuration on the links, $\psi_\sigma\in\mathbb C$ is the ``weight'' of the sign configuration $\sigma$, and the prime indicates that the sum is only over those configurations $\sigma$ that have even number of $-$'s on the links around any site---this condition is enforced by the Gauss law constraint \eqref{gausslawconstraint}.

Consider the action of the operator $P_\ell Q_p$ on the basis state $|\sigma\>$,
\ie
&P_\ell Q_p |\cdots\sigma_{\ell_1}\sigma_{\ell_2}\sigma_{\ell_3}\sigma_{\ell_4}\cdots\>
\\
&= \begin{cases}
0~, & \sigma_\ell = +~,
\\
\frac12(|\cdots\sigma_{\ell_1}\sigma_{\ell_2}\sigma_{\ell_3}\sigma_{\ell_4}\cdots\> - |\cdots\bar\sigma_{\ell_1}\bar\sigma_{\ell_2}\bar\sigma_{\ell_3}\bar\sigma_{\ell_4}\cdots\>)~,&  \sigma_\ell = -~,
\end{cases}
\fe
where $\ell_1,\ell_2,\ell_3,\ell_4$ are the four links around the plaquette $p$ and $\bar\sigma_{\ell_i} = -\sigma_{\ell_i}$. Then, the constraint $P_\ell Q_p |\psi\> = 0$ gives the relation
\ie\label{psi-relation}
\psi_{\cdots\sigma_{\ell_1}\sigma_{\ell_2}\sigma_{\ell_3}\sigma_{\ell_4}\cdots} = \psi_{\cdots\bar\sigma_{\ell_1}\bar\sigma_{\ell_2}\bar\sigma_{\ell_3}\bar\sigma_{\ell_4}\cdots}~,\quad \text{if}\quad \sigma_\ell = -~.
\fe
Note that the constraint $P_\ell Q_p |\psi\> = 0$ does not give any relation when $\sigma_\ell = +$. In particular, there is no relation involving $\psi_{{+}\cdots{+}}$, i.e., all relations are among $\psi_\sigma$'s with $\sigma \ne {+}\cdots{+}$. More importantly, the relation \eqref{psi-relation} is consistent with the Gauss law constraint in the decomposition \eqref{arbitrarystate'}, i.e., if we start with a sign configuration that respects the Gauss law, then flipping all the signs around a plaquette gives another configuration that respects the Gauss law. In addition, such a flip also preserves the charges of the configuration under the $\mathbb Z_2$ 1-form symmetry.

Let us phrase the relation \eqref{psi-relation} between the weights $\psi_\sigma$'s differently. Given a configuration $\sigma \ne {+}\cdots{+}$, pick a link $\ell$ with a $-$ and a plaquette $p\perp \ell$. Consider the configuration $\sigma'$ obtained by flipping all the signs around $p$. Then, $\psi_{\sigma'} = \psi_\sigma$. More generally, whenever $\sigma'$ is obtained from $\sigma$ by a sequence of such flips, we have $\psi_{\sigma'} = \psi_\sigma$.

\begin{figure}
\centering
\includegraphics[scale=0.2]{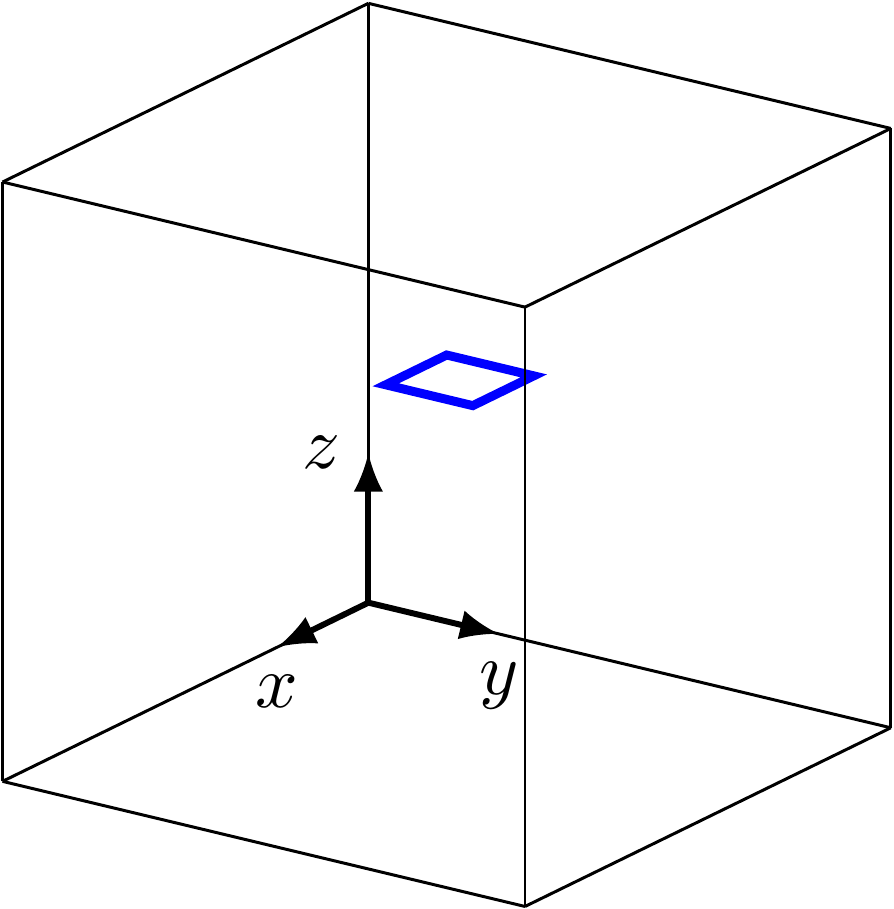} \hfill \includegraphics[scale=0.2]{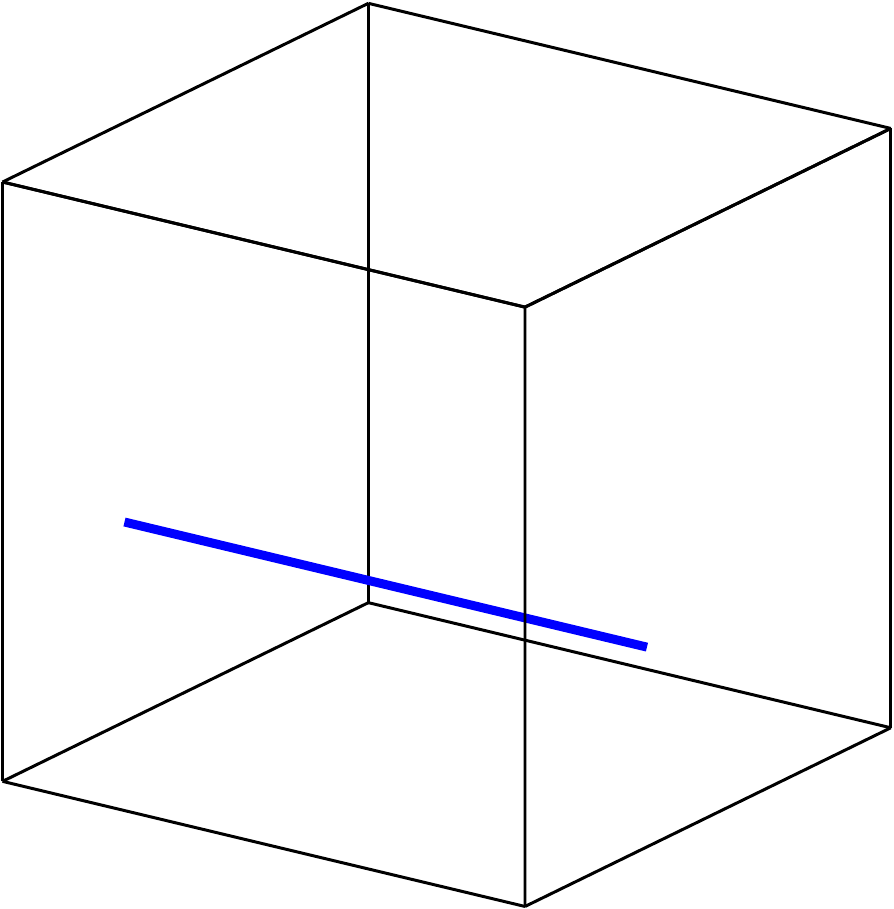} \hfill \includegraphics[scale=0.2]{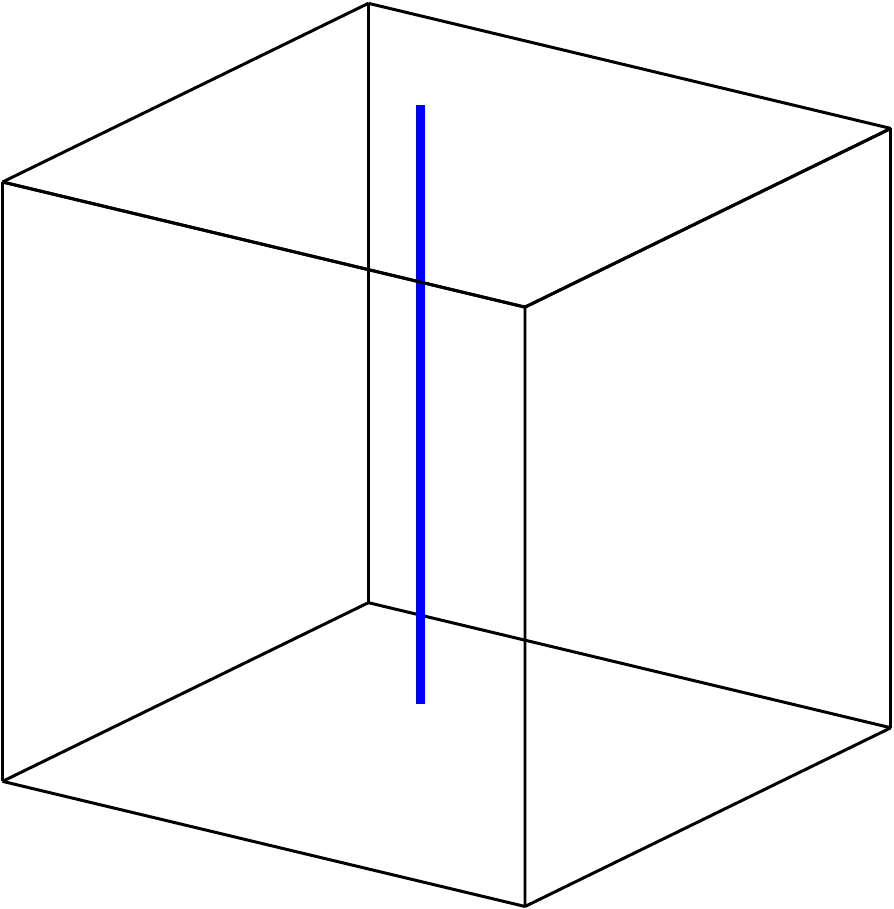} \hfill \includegraphics[scale=0.2]{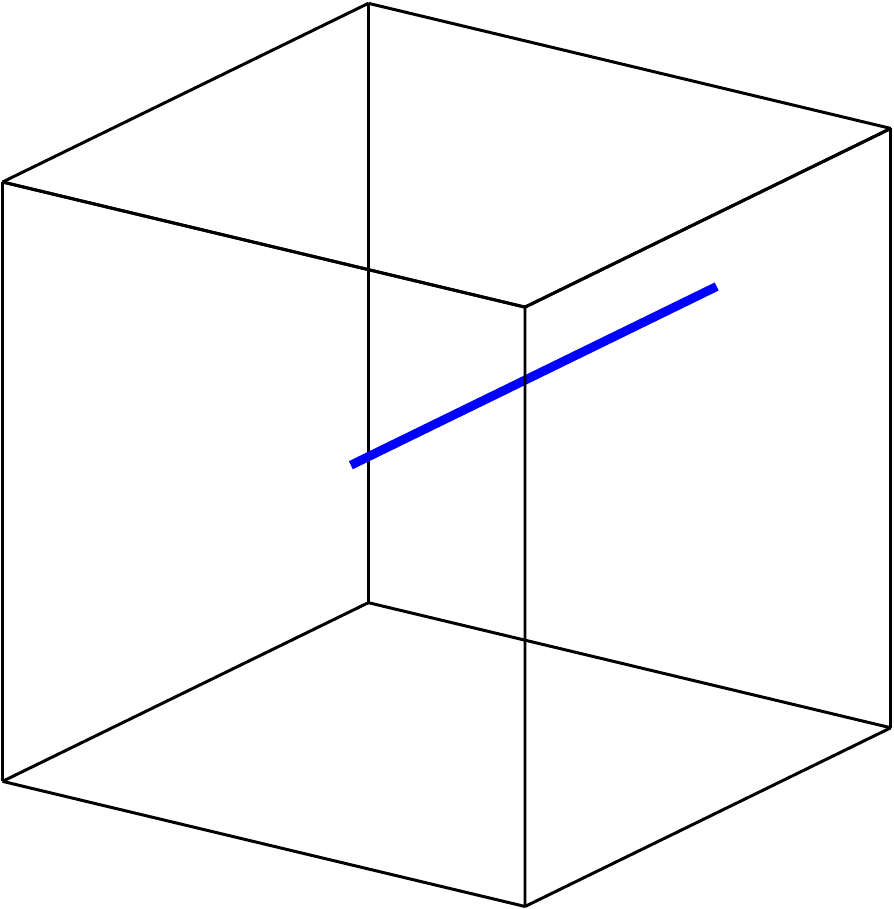}
\\
~~~~(a) $\sigma^{000}$ \hfill (b) $\sigma^{100}$ \hfill (c) $\sigma^{010}$ \hfill (d) $\sigma^{001}$~~~~
\\~~\\
\includegraphics[scale=0.2]{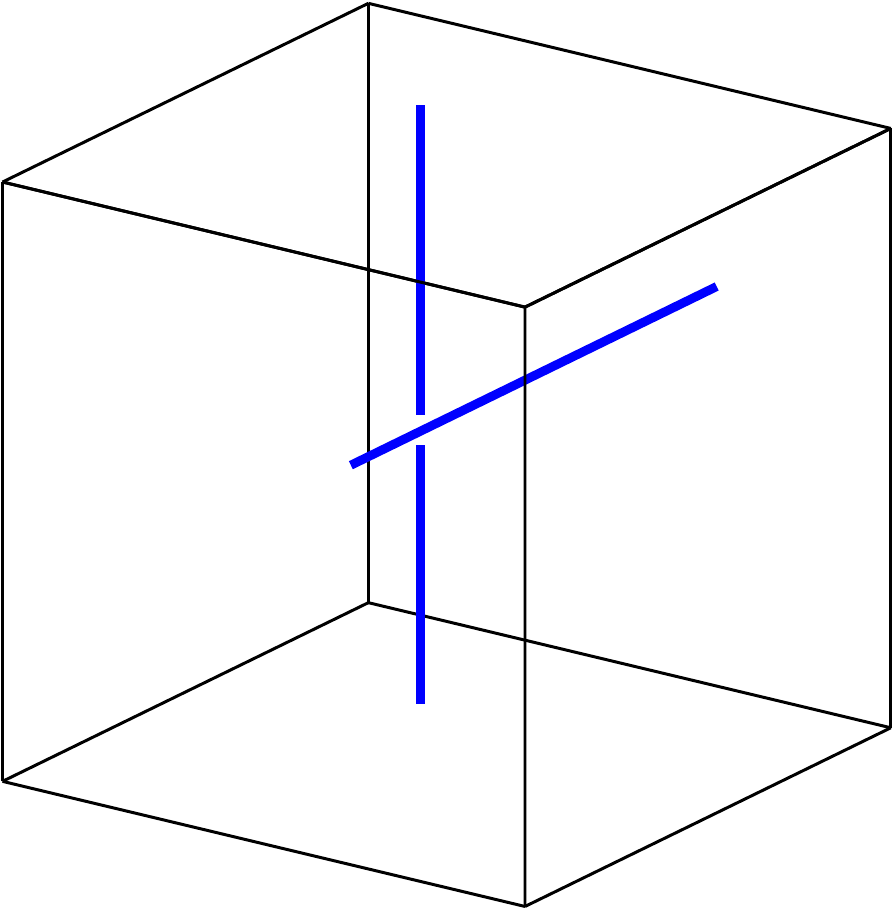} \hfill \includegraphics[scale=0.2]{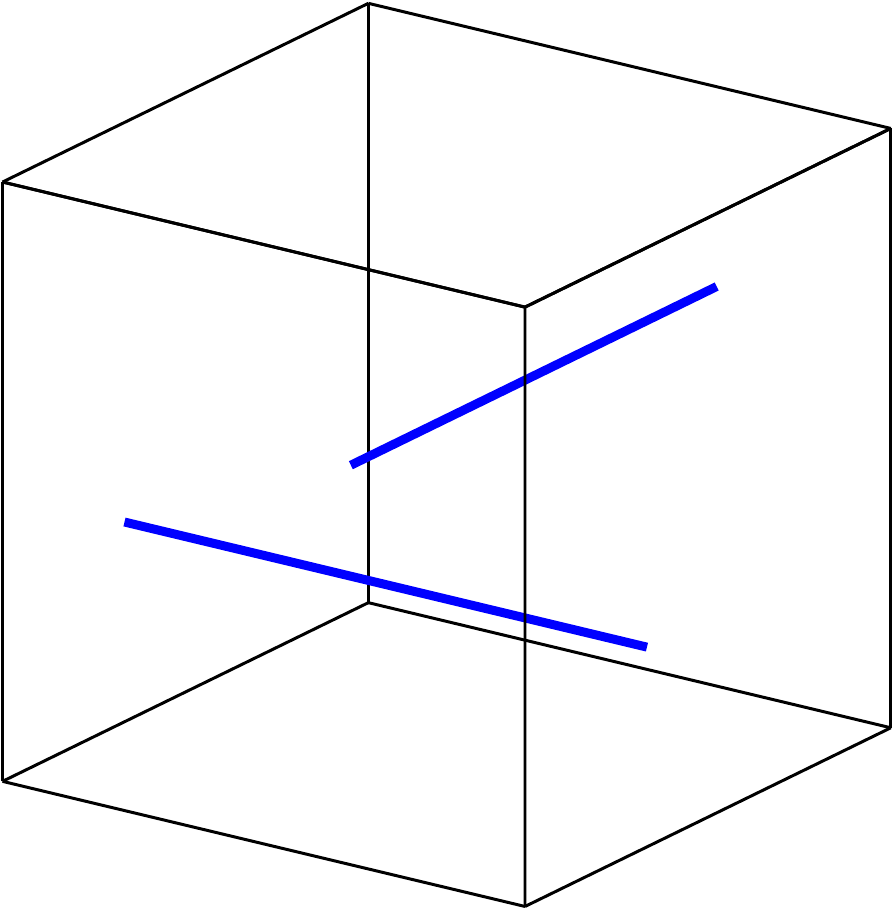} \hfill \includegraphics[scale=0.2]{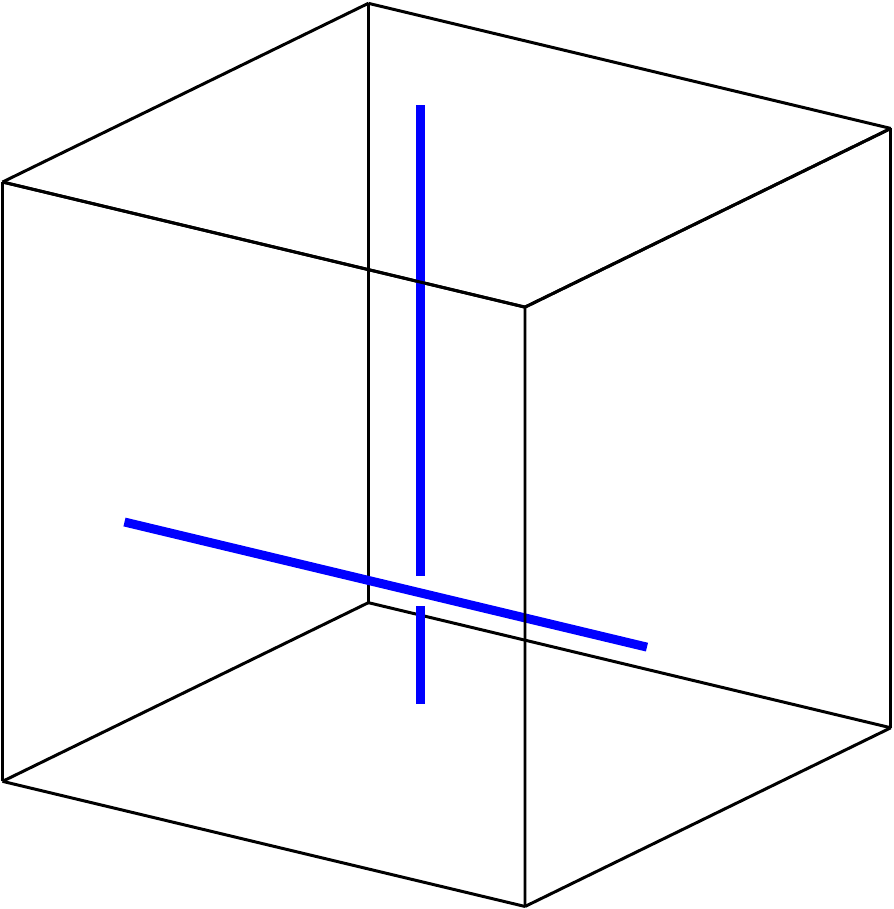} \hfill \includegraphics[scale=0.2]{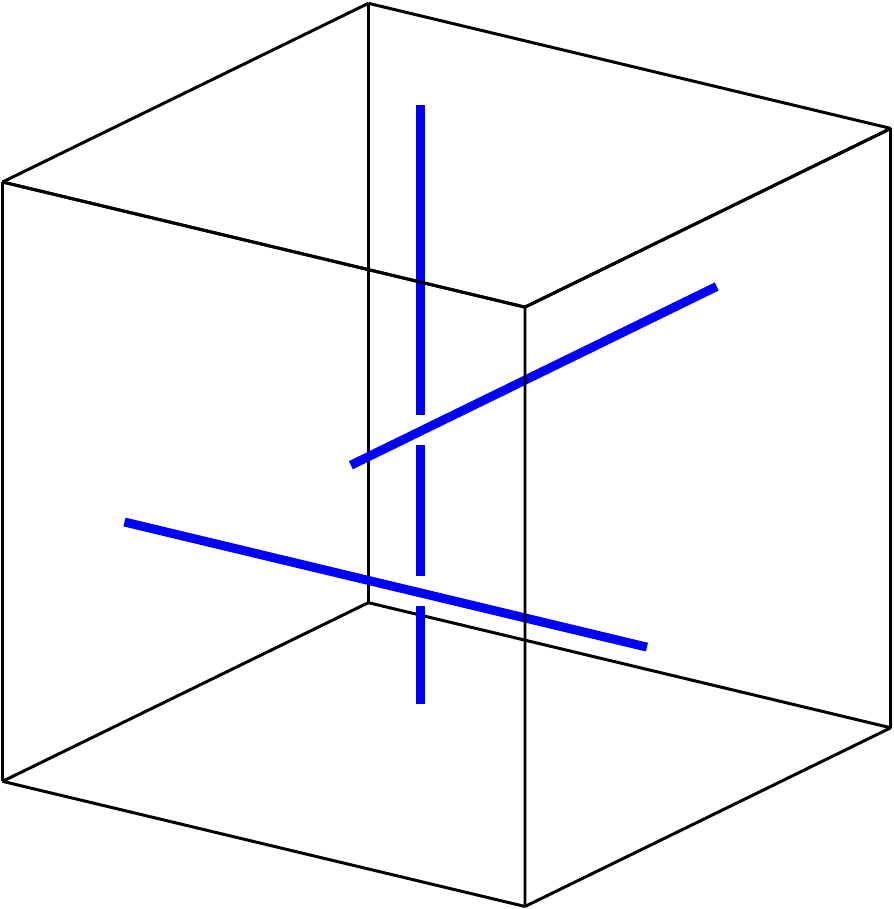}
\\
~~~~(e) $\sigma^{011}$ \hfill (f) $\sigma^{101}$ \hfill (g) $\sigma^{110}$ \hfill (h) $\sigma^{111}$~~~~
\caption{The eight sign configurations associated with the states $|\sigma^\zeta\>$ defined in \eqref{eight-config}. Here, the black box represents the periodic cubic lattice with opposite faces identified (i.e., a torus), the blue links represent links with $-$, and the rest of the links have $+$. Since any flip preserves the $\mathbb Z_2$ 1-form symmetry charges, none of these eight configurations is related to another by a sequence of flips. The non-contractible blue cycles in (b-h) correspond to the insertions of the Wilson line operators in \eqref{eight-config}.}\label{fig:config}
\end{figure}

We claim that any two distinct sign configurations $\sigma \ne {+}\cdots{+}$ and $\sigma' \ne {+}\cdots{+}$ are related by a sequence of flips if and only if they carry the same $\mathbb Z_2^{(1)}$ charges, i.e., $\eta_{ij} |\sigma\> = \eta_{ij} |\sigma'\>$ for all $i<j$. That this is necessary is clear because any flip preserves the $\mathbb Z_2^{(1)}$ charges. To see that this is sufficient, it is enough to show that each $\sigma \ne {+}\cdots{+}$ can be reduced by a sequence of flips to exactly one of the eight configurations associated with the states:
\ie\label{eight-config}
|\sigma^{000}\> := \prod_{\ell \in p_0} Z_\ell |{+}\cdots{+}\>~,\qquad |\sigma^{\zeta\ne 000}\> := \prod_{i,j,k\atop\text{cyclic}} W_k^{\zeta_{ij}} |{+}\cdots{+}\>~,
\fe
where $p_0$ is an arbitrarily chosen plaquette and $\zeta = (\zeta_{xy},\zeta_{yz},\zeta_{zx}) \in \{0,1\}^3$ labels the $\mathbb Z_2^{(1)}$ charges: $\eta_{ij} |\sigma^\zeta\> = (-1)^{\zeta_{ij}} |\sigma^\zeta\>$. These eight configurations are illustrated in Figure \ref{fig:config}. It is not an accident that the eight configurations resemble the eight homology classes of $H_1(T^3,\mathbb Z_2)$, the first homology group of 3-torus.

Before proving the claim, let us establish some useful sequences of flips. Given a configuration $\sigma \ne {+}\cdots{+}$, consider a plaquette $p$.
\begin{enumerate}
\item If there is a link $\ell$ such that $\ell \perp p$ and $\sigma_\ell = -$, then we can flip all the signs around $p$.
\item Else, if there is a link $\ell$ such that $\ell$ meets $p$ at a site, $\ell$ is in the plane of $p$, and $\sigma_\ell = -$, then we can flip all the signs around $p$ using the following sequence of flips:
\ie\label{eq:useful-flip-seq}
\raisebox{-0.5\totalheight}{\includegraphics[scale=0.2]{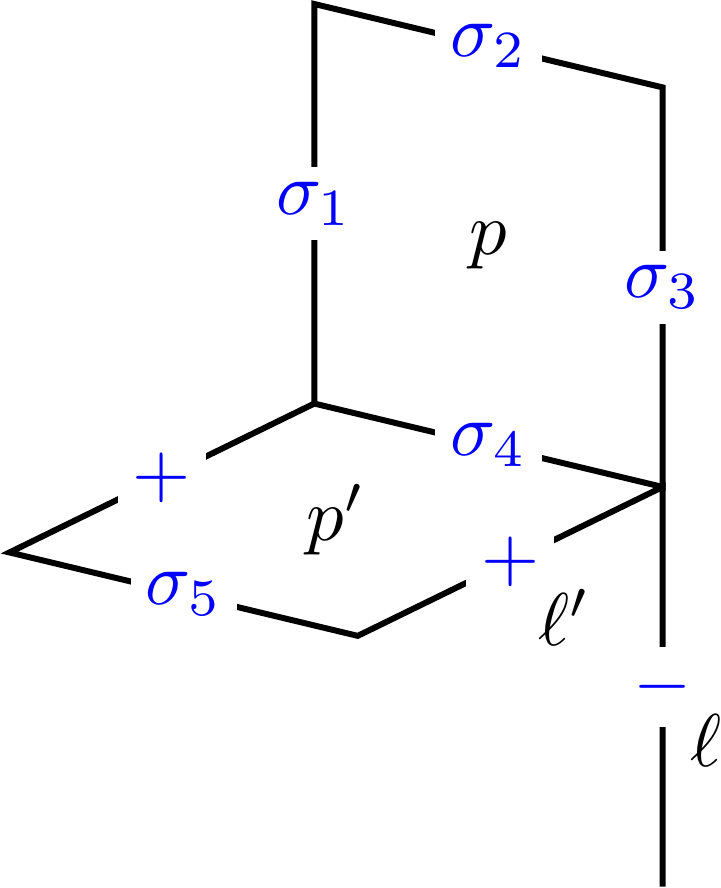}}~\overset{P_\ell Q_{p'}}\longleftrightarrow~\raisebox{-0.5\totalheight}{\includegraphics[scale=0.2]{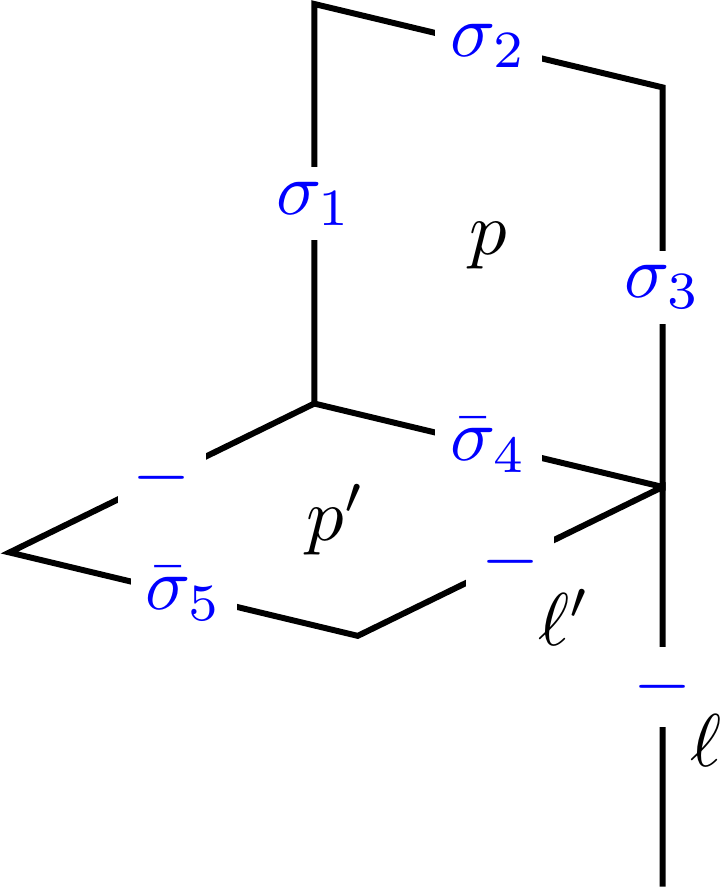}}~\overset{P_{\ell'}Q_p}\longleftrightarrow~\raisebox{-0.5\totalheight}{\includegraphics[scale=0.2]{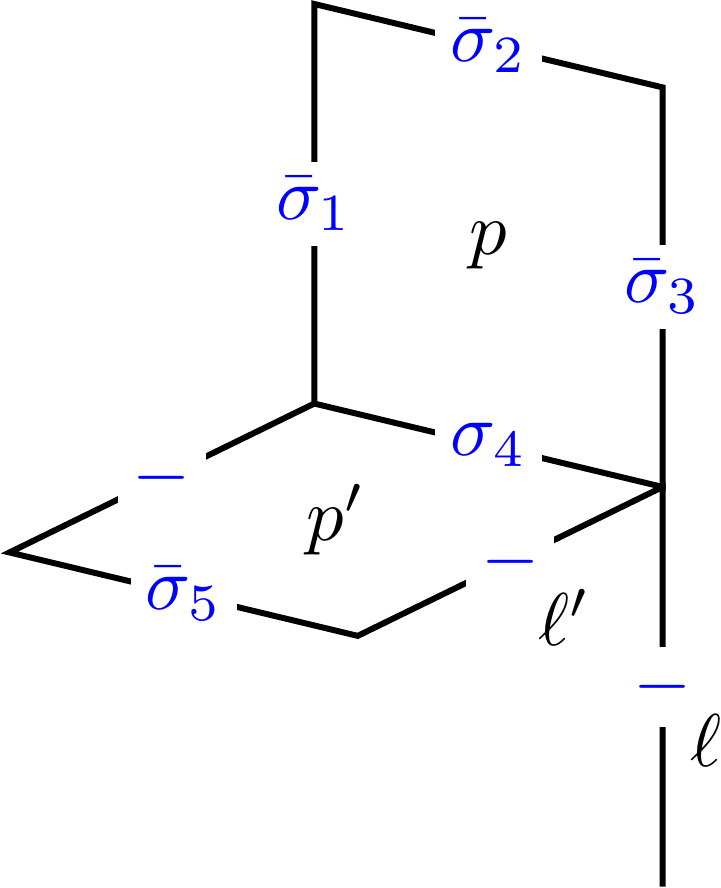}}~\overset{P_\ell Q_{p'}}\longleftrightarrow~\raisebox{-0.5\totalheight}{\includegraphics[scale=0.2]{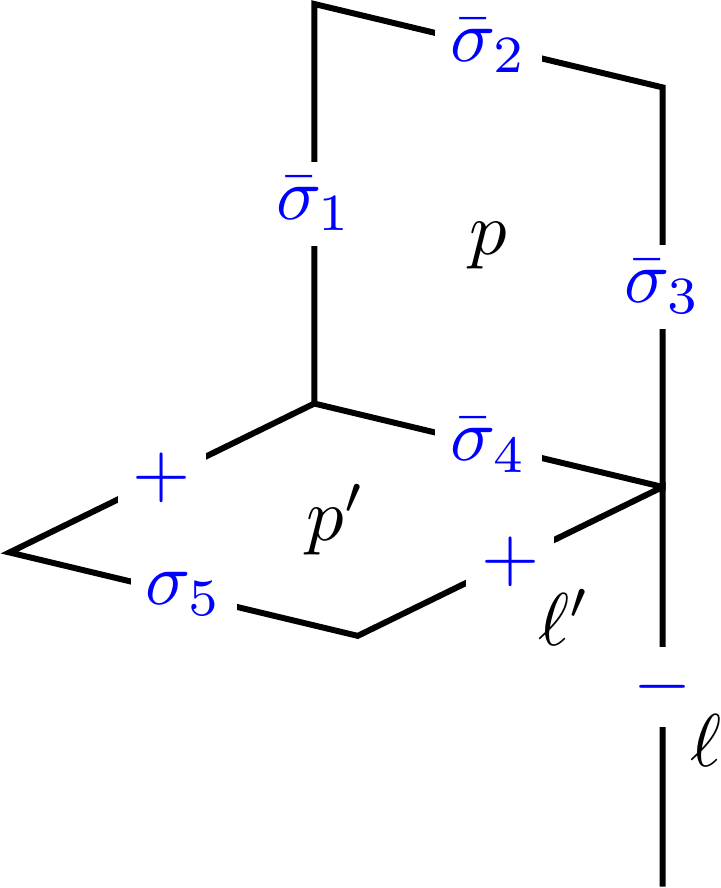}}~.
\fe
\item Else, if there is a link $\ell$ such that $\ell$ and $p$ belong to the same cube, $\ell$ does not meet $p$, and $\sigma_\ell = -$, then we can flip all the signs around $p$ using the following sequence of flips:
\ie\label{eq:useful-flip1-seq}
\raisebox{-0.5\totalheight}{\includegraphics[scale=0.2]{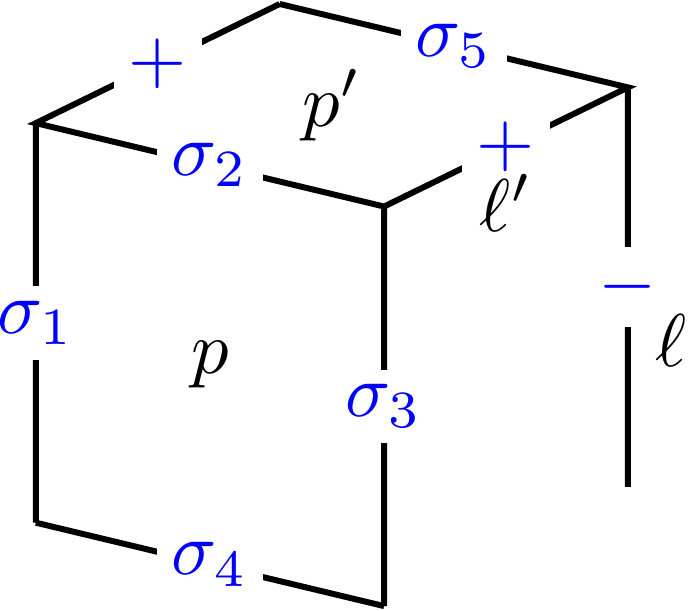}}~\overset{P_\ell Q_{p'}}\longleftrightarrow~\raisebox{-0.5\totalheight}{\includegraphics[scale=0.2]{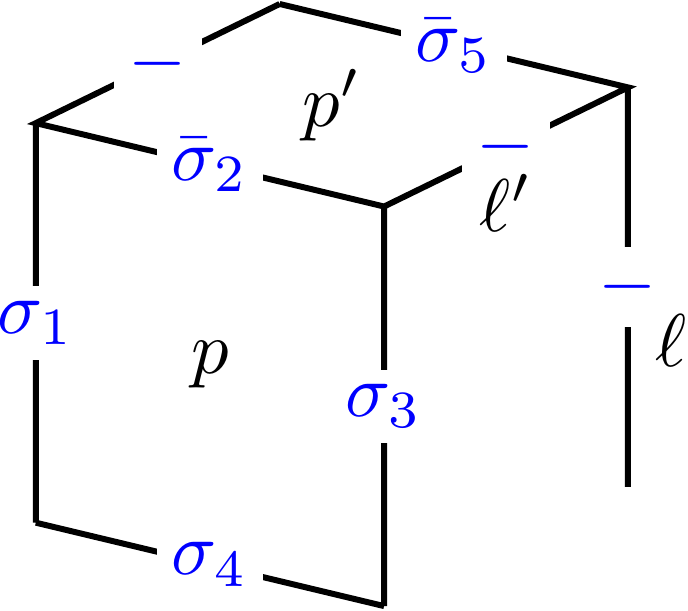}}~\overset{P_{\ell'}Q_p}\longleftrightarrow~\raisebox{-0.5\totalheight}{\includegraphics[scale=0.2]{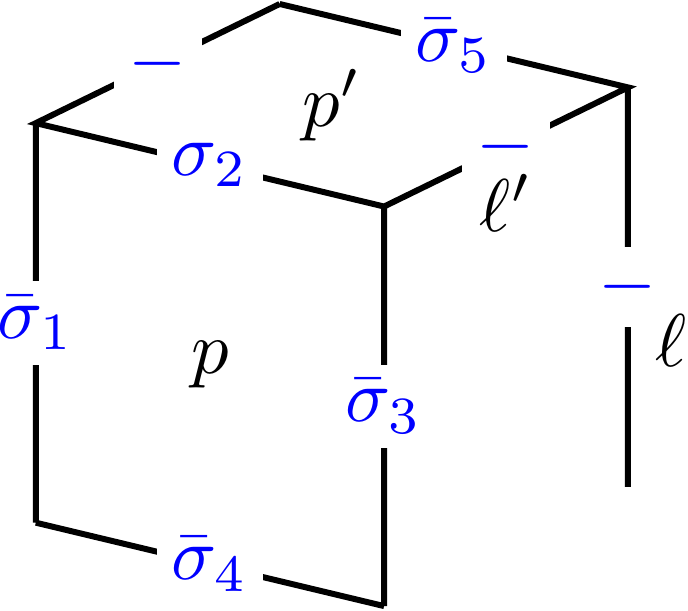}}~\overset{P_\ell Q_{p'}}\longleftrightarrow~\raisebox{-0.5\totalheight}{\includegraphics[scale=0.2]{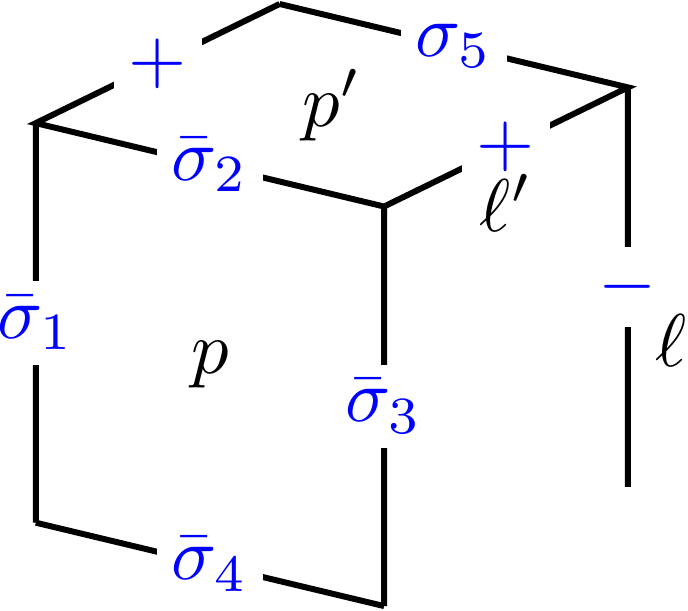}}~.
\fe
\item Else, if not all the signs around $p$ are the same, then we can flip all the signs around $p$ using one of the following sequences of flips:
\ie\label{eq:useful-flip2-seq}
\raisebox{-0.5\totalheight}{\includegraphics[scale=0.2]{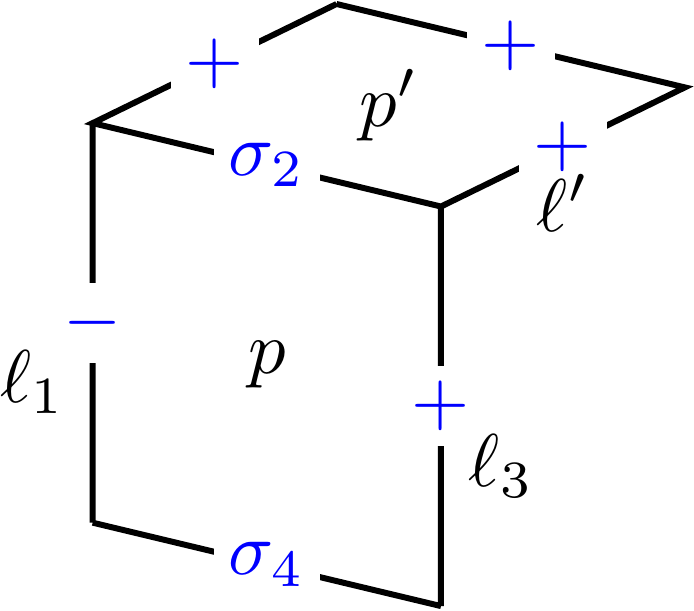}}~\overset{P_{\ell_1} Q_{p'}}\longleftrightarrow~\raisebox{-0.5\totalheight}{\includegraphics[scale=0.2]{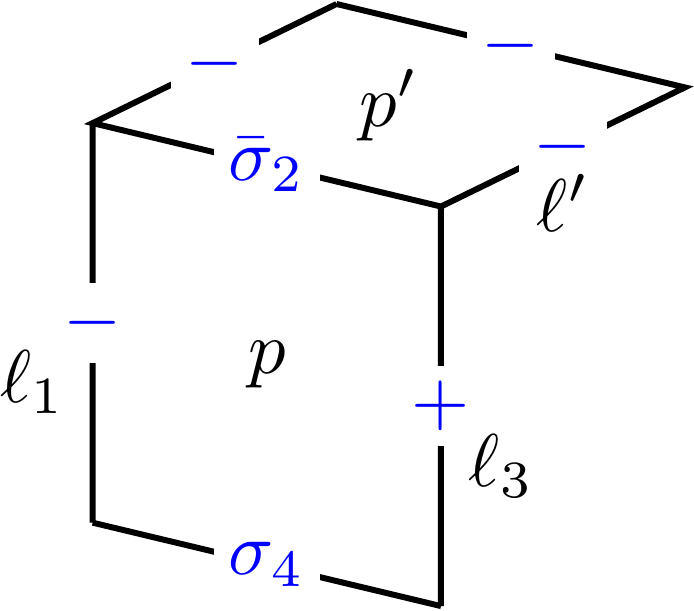}}~\overset{P_{\ell'}Q_p}\longleftrightarrow~\raisebox{-0.5\totalheight}{\includegraphics[scale=0.2]{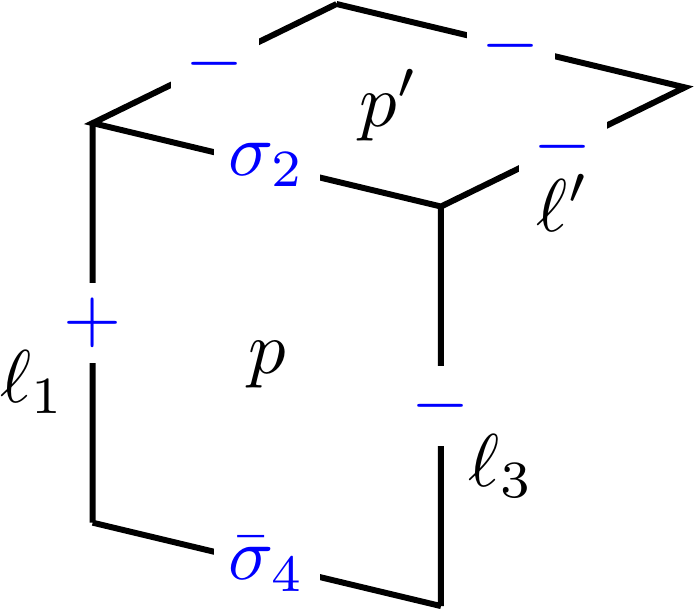}}~\overset{P_{\ell_3} Q_{p'}}\longleftrightarrow~\raisebox{-0.5\totalheight}{\includegraphics[scale=0.2]{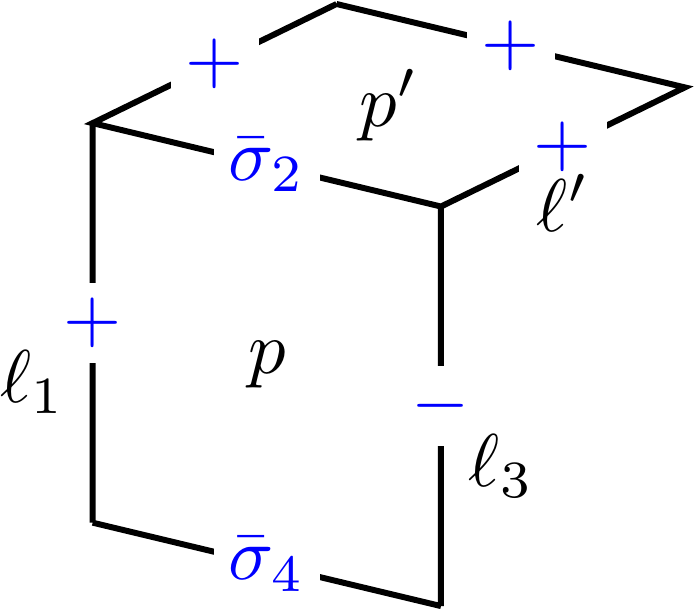}}~,
\fe
or
\ie\label{eq:useful-flip3-seq}
\raisebox{-0.5\totalheight}{\includegraphics[scale=0.2]{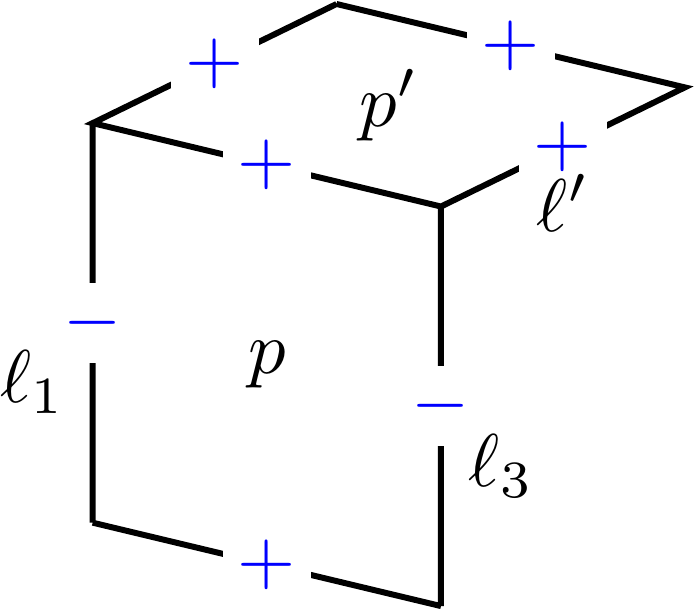}}~\overset{P_{\ell_1} Q_{p'}}\longleftrightarrow~\raisebox{-0.5\totalheight}{\includegraphics[scale=0.2]{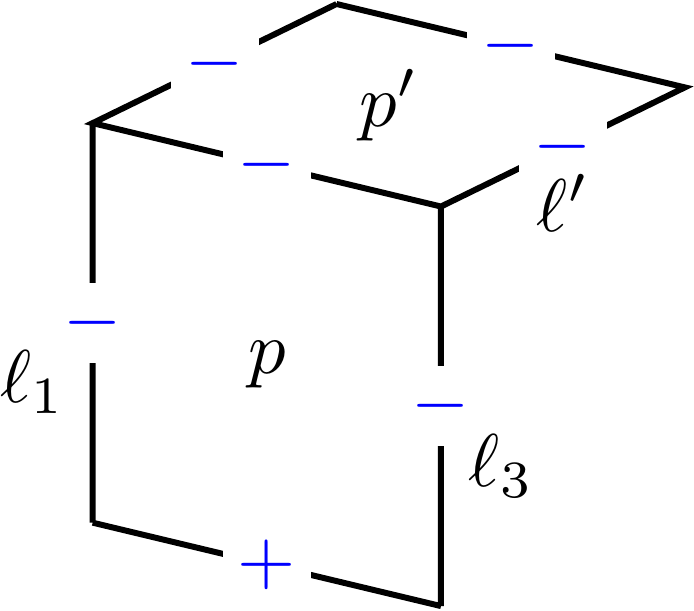}}~\overset{P_{\ell'}Q_p}\longleftrightarrow~\raisebox{-0.5\totalheight}{\includegraphics[scale=0.2]{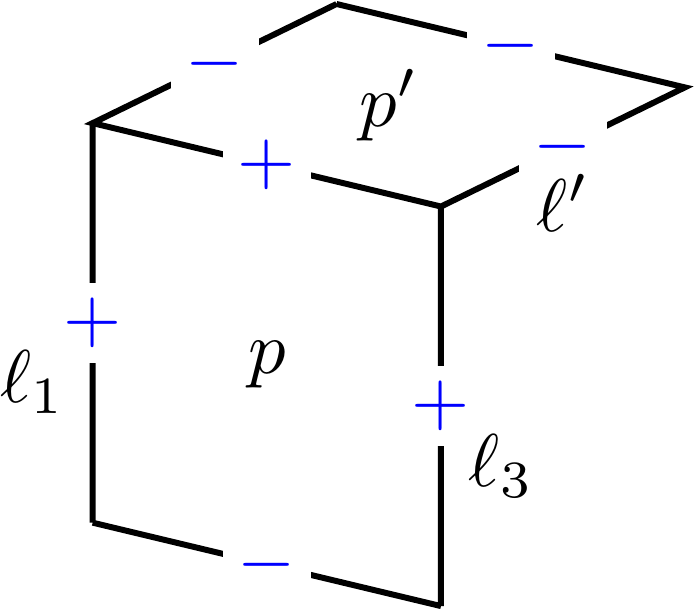}}~\overset{\text{flip }p'}\longleftrightarrow~\raisebox{-0.5\totalheight}{\includegraphics[scale=0.2]{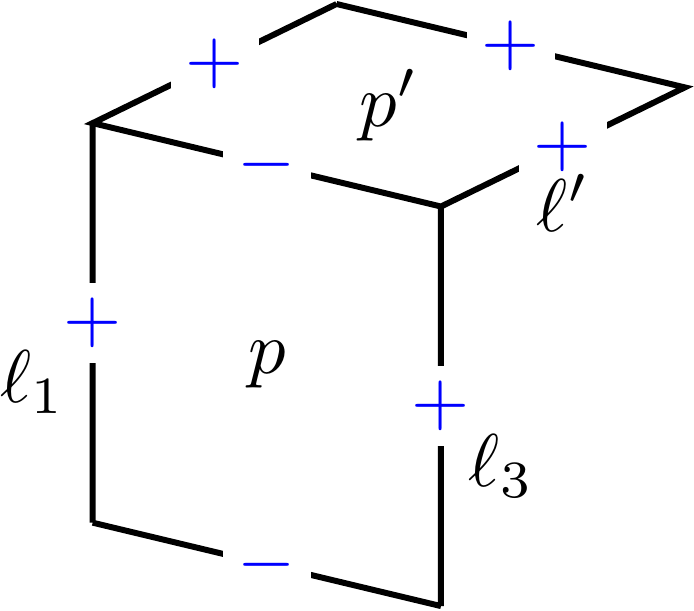}}~,
\fe
where the last step uses one of the previous sequences of flips to flip the signs around the plaquette $p'$.
\end{enumerate}
(Note that in sequences 1, 3, and 4, all the flips are performed within a single cube.) In conclusion, in any configuration $\sigma\ne {+}\cdots{+}$, all the signs around any plaquette $p$ can be flipped without affecting any other signs, except when the signs around $p$ are all the same and the signs on all the links near $p$ are all $+$.

Now, as in Figure \ref{fig:config}, consider a pictorial representation of the sign configurations where links with $-$ are colored blue and links with $+$ are uncolored.\footnote{This color scheme is consistent with the usage of blue links for the Pauli $Z$ operators in Figures \ref{fig:Hterms} and \ref{fig:deformHterm}, and the fact that $|-\> = Z |+\>$.} Then, the Gauss law constraints imply that every site has an even number of blue links around it. Therefore, in any configuration $\sigma$, the blue links form closed cycles on the lattice. Using the sequences of flips described in the last paragraph, we can perform the following ``moves'' on any plaquette:
\ie\label{3d-moves}
\text{i: }\quad&\raisebox{-0.5\totalheight}{\includegraphics[scale=0.3]{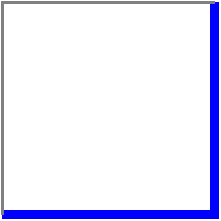}} \quad\longleftrightarrow\quad \raisebox{-0.5\totalheight}{\includegraphics[scale=0.3]{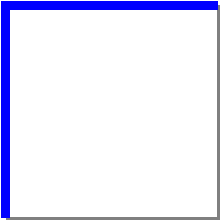}}~,\qquad\qquad &\text{ii: }\quad&\raisebox{-0.5\totalheight}{\includegraphics[scale=0.3]{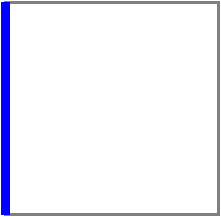}} \quad\longleftrightarrow\quad \raisebox{-0.5\totalheight}{\includegraphics[scale=0.3]{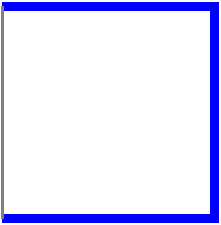}}~,
\\~~\\
\text{iii: }\quad&\raisebox{-0.5\totalheight}{\includegraphics[scale=0.3]{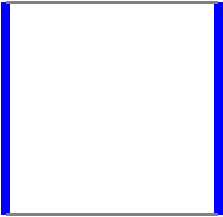}} \quad\longleftrightarrow\quad \raisebox{-0.5\totalheight}{\includegraphics[scale=0.3]{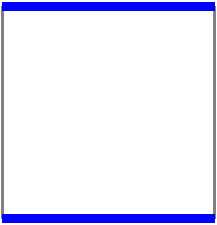}}~,\qquad\qquad &\text{iv: }\quad&\raisebox{-0.5\totalheight}{\includegraphics[scale=0.3]{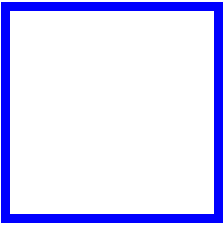}} \quad\overset{?}\longleftrightarrow\quad \raisebox{-0.5\totalheight}{\includegraphics[scale=0.3]{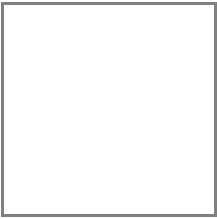}}~.
\fe
On one hand, moves i, ii, and iii are always possible because we can perform at least one of the four sequences of flips described in the last paragraph. On the other hand, move iv is not always possible (hence the $?$). In fact, it is not possible exactly when all the links near the plaquette have $+$. In other words, move iv is not possible exactly when the blue loop is disconnected/isolated from any other blue cycle.

These moves are reminiscent of moves one can perform on a cycle to obtain another cycle that is homologous to it. The only difference is that we cannot shrink the isolated blue loops around a plaquette to a point. It is well-known from homology that, using these moves, any configuration of cycles can be ``almost'' reduced to exactly one of the eight configurations in Figure \ref{fig:config}. We say ``almost'' because we are still left with isolated blue loops. Let us deal with them now.

We can translate an isolated blue loop in its plane:
\ie\label{translate}
\text{translate:}\quad\raisebox{-0.5\totalheight}{\includegraphics[scale=0.3]{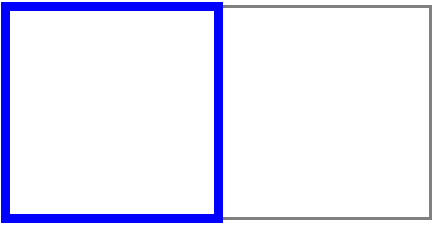}} \quad\longrightarrow\quad \raisebox{-0.5\totalheight}{\includegraphics[scale=0.3]{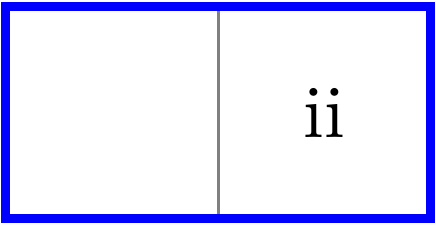}} \quad\longrightarrow\quad \raisebox{-0.5\totalheight}{\includegraphics[scale=0.3]{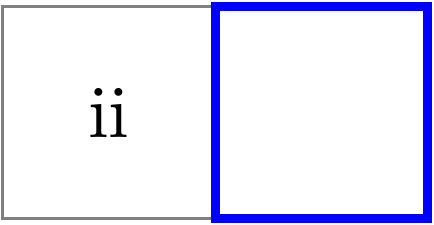}}~,
\fe
where the roman numeral and its position indicate which move is performed on which plaquette. We can also change its orientation:
\ie\label{rotate}
\text{rotate:}\quad\raisebox{-0.5\totalheight}{\includegraphics[scale=0.3]{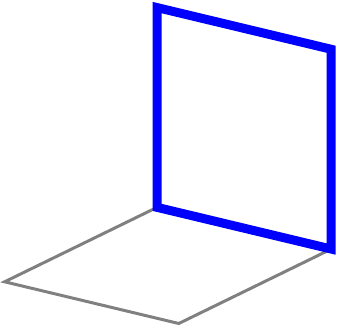}} \quad\longrightarrow\quad \raisebox{-0.5\totalheight}{\includegraphics[scale=0.3]{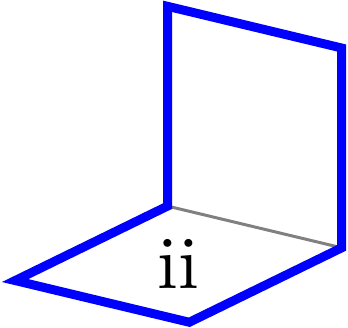}} \quad\longrightarrow\quad \raisebox{-0.5\totalheight}{\includegraphics[scale=0.3]{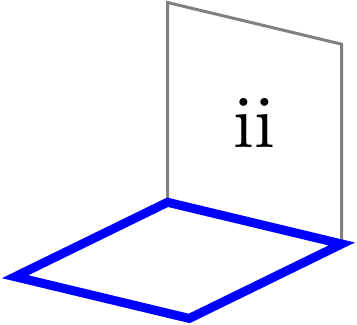}}~.
\fe
Combining such moves, any isolated blue loop can be translated and rotated arbitrarily provided we do not run into other blue links during the move. If we do run into a blue link, we can absorb the isolated blue loop into that blue link:
\ie\label{absorb}
\text{absorb:}\quad\raisebox{-0.5\totalheight}{\includegraphics[scale=0.3]{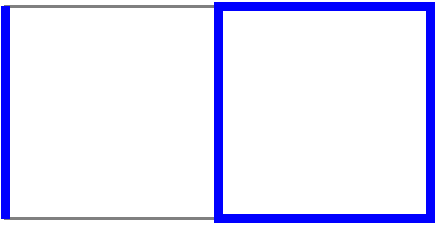}} ~\longrightarrow~ \raisebox{-0.5\totalheight}{\includegraphics[scale=0.3]{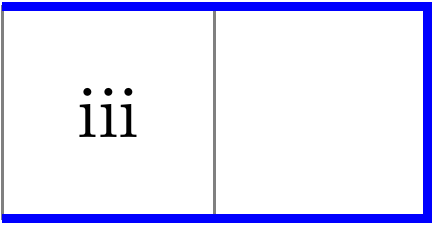}}~\longrightarrow~ \raisebox{-0.5\totalheight}{\includegraphics[scale=0.3]{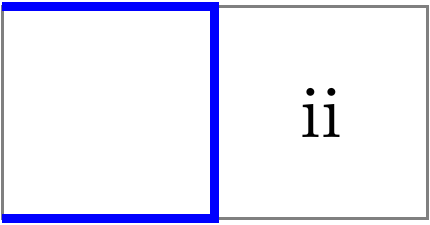}} ~\longrightarrow~ \raisebox{-0.5\totalheight}{\includegraphics[scale=0.3]{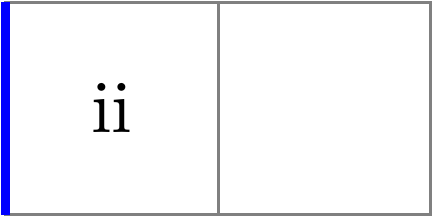}}~.
\fe
Note that we used only moves ii and iii, but not iv, so these moves are always possible.

This way, we can reduce any number of isolated blue loops to a single isolated blue loop. This produces the configuration $\sigma^{000}$ in Figure \ref{fig:config}(a) in the absence of non-contractible cycles. If there is a non-contractible cycle, we can further absorb the remaining isolated blue loop into that cycle. This produces the rest of the configurations in Figure \ref{fig:config}. This completes the proof of the claim that any $\sigma\ne {+}\cdots{+}$ is related by a sequence of flips to exactly one of the eight configurations in \eqref{eight-config}.

It follows that $\psi_\sigma = \psi_{\sigma^\zeta}$ for all $\sigma \ne {+}\cdots{+}$ that satisfy $\eta_{ij}|\sigma\> = (-1)^{\zeta_{ij}} |\sigma\>$. Hence, $|\psi\>$ can be written as
\ie
|\psi\> = \alpha |{+}\cdots{+}\> + \sum_{\zeta\in\{0,1\}^3} \psi_{\sigma^\zeta} \sideset{}{'}\sum_{\sigma\in\{+,-\}^{3V}\atop\eta_{ij}|\sigma\>=(-1)^{\zeta_{ij}}|\sigma\>} |\sigma\>~,
\fe
where $\alpha := \psi_{{+}\cdots{+}} - \psi_{\sigma^{000}}$. Up to scalar factors, we have\footnote{The first equality follows from the claim before that any two sign configurations are related by a sequence flips if and only if they have the same $\mathbb Z_2^{(1)}$ charges. The second equality follows from \eqref{zeta-loopgas}.}
\ie
\sideset{}{'}\sum_{\sigma\in\{+,-\}^{3V}\atop\eta_{ij}|\sigma\>=(-1)^{\zeta_{ij}}|\sigma\>} |\sigma\> \propto \prod_p \left( \frac{1+\prod_{\ell \in p} Z_\ell}{2} \right) |\sigma^\zeta\> \propto |\zeta\>~,
\fe
so we can equivalently write $|\psi\>$ as
\ie
|\psi\> = \alpha |{+}\cdots{+}\> + \sum_{\zeta\in\{0,1\}^3} \alpha_\zeta |\zeta\>~,
\fe
for some $\alpha_\zeta \in \mathbb C$. Changing to the $|\xi\>$ basis, we can also write it as 
\ie
|\psi\> = \alpha |{+}\cdots{+}\> + \sum_{\xi\in\{0,1\}^3} \beta_\xi |\xi\>~,
\fe
for some $\beta_\xi \in \mathbb C$. This shows that the nine states in \eqref{groundstates} are the only ground states of $\widetilde H'$.

\subsection{Proof of gap}\label{sec:3d-proofofgap-martingale}
In this subsection, we use the martingale method \cite{Fannes:1992,Nachtergaele:1996,Kastoryano_2018} to prove that $\widetilde H'$ \eqref{3d-deformedtH'} is gapped in the thermodynamic limit.

In general, proving the gap of an arbitrary local Hamiltonian is a very hard problem \cite{Cubitt:2015xsa,Cubitt:2015lta}. For frustration-free Hamiltonians, there are two well-known methods to prove the gap: the Knabe method \cite{Knabe:1988,Gosset:2016}, and the martingale method \cite{Fannes:1992,Nachtergaele:1996,Kastoryano_2018}. Both methods ultimately rely on Knabe's argument \cite{Knabe:1988}: if $\Delta$ denotes the gap of a positive semi-definite operator $H$, then, for any $\Delta_0 \ge 0$, $H^2 \ge \Delta_0 H \implies \Delta \ge \Delta_0$. So we just have to find a $\Delta_0 > 0$, independent of system size, such that $H^2 \ge \Delta_0 H$. Where the two methods differ is in how $\Delta_0$ is computed. The Knabe method relies on numerical techniques to compute $\Delta_0$, which is feasible in 1+1d, but impractical in higher dimensions. In contrast, the martingale method relates $\Delta_0$ to a quantity associated with ``local ground states''. (See Appendix \ref{app:delta} for a discussion on this quantity.) So the latter is entirely analytical, provided the ``local ground states'' are known explicitly. Since we know the ground states of our model explicitly, we use the martingale method below.

It is conceptually simpler to apply this method on a tensor product Hilbert space, so instead of imposing the Gauss law constraint exactly, we impose it energetically like in \eqref{H}. In this case, the total Hilbert space is the tensor product Hilbert space $\mathcal H$. The deformation preserving the (non-topological) $\mathbb Z_2$ 1-form symmetry and the Wegner duality symmetry would then be
\ie
H_\lambda = -J \left( \sum_p \prod_{\ell\in p}Z_\ell + \frac{h}{J}\sum_\ell X_\ell + \frac{\lambda}{8} \sum_{\ell,p:\ell\perp p} X_\ell \prod_{\ell'\in p} Z_{\ell'} \right) - g \sum_s G_s~.
\fe
Once again, we set $J=h$ and $\lambda = 1$, and consider a shifted Hamiltonian
\ie\label{3d-deformedH'}
H' = \frac{J}{2} \sum_{\ell,p:\ell\perp p} P_\ell Q_p + g \sum_s (1-G_s)~,
\fe
which differs from $H_{\lambda=1}$ by only a scalar multiple of the identity operator.

When $g>0$, the ground states of $H'$ are the same as those of $\widetilde H'$. Moreover, the gap of $H'$ is at most the gap of $\widetilde H'$ for any $g>0$, with equality when $g$ is large enough. Therefore, it is enough to show that $H'$ is gapped in the thermodynamic limit for any $g>0$.

Let us write $H' = H'' + H'''$, where
\ie\label{3d-deformedH''}
H'' = \frac{J}{2} \sum_{\ell,p:\ell\perp p} P_\ell Q_p~,\qquad H''' = g \sum_s (1-G_s)~.
\fe
While $H''$ is the same as $\widetilde H'$, they act on different Hilbert spaces: $H''$ acts on the tensor product Hilbert space $\mathcal H$, whereas $\widetilde H'$ acts on the constrained Hilbert space $\widetilde{\mathcal H}$.

Since $H''$ and $H'''$ commute with each other, the eigenvalues of $H'$ are sums of eigenvalues of $H''$ and $H'''$. Therefore, if $H''$ and $H'''$ are gapped then $H'$ is also gapped. All terms in $H'''$ commute with each other, so we can solve for the spectrum of $H'''$ exactly. In particular, it has a gap of $4g>0$ (since $\prod_s G_s = 1$, one can violate only an even number of terms). So it is sufficient to prove that $H''$ is gapped. The ground states of $H''$ are analyzed in Appendix \ref{app:deformed-nogausslaw}.

For simplicity, we assume that $L_i$'s are multiples of an integer $n>0$, which we will fix later. We further assume that each $L_i \ge 3n$. (The following proof can be generalized to other values of $L_i$'s.) Given an $\b x = (x,y,z) \in \mathbb Z^3$, let $C_{\b x}$ be the cube with corners $\{\b x + (2n-1) \b\varepsilon: \b\varepsilon \in \{0,1\}^3\}$. In other words, $C_{\b x}$ is the cube with side length $2n-1$ in the positive octant of the corner $\b x$.

Consider the cubes $C_{n \b a}$ where $a_i = 1,\ldots, L_i/n$ for $i=x,y,z$. These bigger cubes cover all the cubes $c$ of the lattice. In fact, every term in the Hamiltonian $H''$ belongs to at least one and at most eight of these bigger cubes. Therefore, we have 
\ie
H''_L \ge \frac18 \cdot  \frac{J}{2} \sum_{\b a} h_{\b a}~,\qquad \text{where} \qquad h_{\b a} := \sum_{\ell,p\in C_{n \b a}:\ell\perp p} P_\ell Q_p~,
\fe
and we introduced the subscript $L$ (shorthand for $L_i$'s) on $H''$ for clarity. The local term $h_{\b a}$ acts on the local Hilbert space $\mathcal H_{\b a} := \bigotimes_{\ell \in C_{n\b a}} \mathcal H_\ell$ of dimension $2^{3(2n-1)(2n)^2}$, where $3(2n-1)(2n)^2$ is the number of links in the cube $C_{n \b a}$. 

Due to frustration-freeness, $h_{\b a}$ has a nontrivial kernel, i.e., ground states with zero energy.\footnote{The ground-state-space (or kernel) of $h_{\b a}$ is spanned by the states
\ie
|{+}\cdots{+}\>_{C_{n\b a}}~,\qquad |S\>_{C_{n\b a}}:= \prod_{s\in R} G_{s|C_{n\b a}} |0\cdots0\>_{C_{n\b a}}~,
\fe
where $S$ is a subset of sites in $C_{n\b a}$ and
\ie
G_{s|C_{n\b a}} := \prod_{\ell\in C_{n\b a}:\ell\ni s} X_\ell~,
\fe
is the restriction of the Gauss law operator to the cube $C_{n\b a}$. Since $|S\>_{C_{n\b a}} = |C_{n\b a}\smallsetminus S\>_{C_{n\b a}}$ due to the relation $\prod_{s\in C_{n\b a}} G_{s|C_{n\b a}} = 1$, the dimension of the kernel of $h_{\b a}$ is $1+\frac12 \cdot 2^{(2n)^3}$, where $(2n)^3$ is the number of sites in the cube $C_{n\b a}$. It can be shown that these are the only ground states of $h_{\b a}$ using an argument similar to the one in Appendix \ref{app:deformed-nogausslaw}.} Let $ \Pi_{\b a}$ be the projector onto the kernel of $h_{\b a}$ and define $ \Pi_{\b a}^\perp:=1- \Pi_{\b a}$. Let $\epsilon_n>0$ be the gap of $h_{\b a}$, which is independent of $\b a$ due to translation invariance, but can depend on $n$. Then, $h_{\b a} \ge \epsilon_n  \Pi_{\b a}^\perp$, and hence,
\ie
H''_L \ge \frac{J\epsilon_n}{16} \bar H_L~,\qquad \text{where} \qquad \bar H_L := \sum_{\b a}  \Pi_{\b a}^\perp~.
\fe
Frustration-freeness implies that the ground states of $H''_L$ are the same as those of $\bar H_L$, so it suffices to show that $\bar H_L$ is gapped.

Following Knabe's argument \cite{Knabe:1988}, our goal is to show that there is a constant $\Delta_0>0$ independent of $L$ such that $\bar H_L^2 \ge \Delta_0 \bar H_L$ for all sufficiently large $L$. Since $\bar H_L$ is positive semi-definite, it follows that $\bar \Delta_L \ge \Delta_0$ for all sufficiently large $L$, thereby proving that $\bar H_L$ is gapped, as long as $n$ is finite (so that $\epsilon_n > 0$).

Consider the square of the Hamiltonian
\ie
\bar H_L^2 = \sum_{\b a}  \Pi_{\b a}^\perp + \sum_{\b a,\b b: \Vert \b a - \b b\Vert_{\infty} = 1} \{  \Pi_{\b a}^\perp, \Pi_{\b b}^\perp \} + \sum_{\b a,\b b:\Vert \b a - \b b\Vert_{\infty} > 1} \{  \Pi_{\b a}^\perp, \Pi_{\b b}^\perp \}~,
\fe
where $\Vert \b a - \b b\Vert_{\infty} := \max_i |a_i - b_i|$ is the $\ell_\infty$-norm. The first term on the right hand side is simply $\bar H_L$. Note that $\Pi_{\b a}^\perp$ commutes with $\Pi_{\b b}^\perp$ if and only if $\Vert \b a - \b b\Vert_{\infty} \ne 1$. (This follows from that fact that $C_{n\b a}$ and $C_{n\b b}$ do not intersect if $\Vert \b a - \b b\Vert_{\infty} > 1$.) Therefore, the third term on the right hand side is positive semi-definite, but the second term is not necessarily positive semi-definite.

We take care of second term using the martingale method, which relies on the following mathematical facts \cite[Lemma 6.3]{Fannes:1992}: any two orthogonal projections $ \Pi_1$ and $ \Pi_2$ satisfy
\ie
&\{ \Pi_1, \Pi_2\} \ge - \Vert  \Pi_1  \Pi_2 -  \Pi_1 \wedge  \Pi_2 \Vert~( \Pi_1 +  \Pi_2)~,
\fe
and
\ie
&\Vert  \Pi_1  \Pi_2 -  \Pi_1 \wedge  \Pi_2 \Vert = \Vert  \Pi_1^\perp  \Pi_2^\perp -  \Pi_1^\perp \wedge  \Pi_2^\perp \Vert~,
\fe
where $\Vert \cdot \Vert$ denotes the operator norm, $ \Pi_1 \wedge  \Pi_2$ is the orthogonal projection onto $\im( \Pi_1) \cap \im( \Pi_2)$, and $ \Pi_{1,2}^\perp := 1- \Pi_{1,2}$. In our case, with $ \Pi_1 =  \Pi_{\b a}^\perp$ and $ \Pi_2 =  \Pi_{\b b}^\perp$, the inequality is\footnote{This inequality is known as the ``martingale condition'', which inspired the name ``martingale method''.}
\ie
\{ \Pi_{\b a}^\perp, \Pi_{\b b}^\perp\} \ge - \delta_n(\b a,\b b) ( \Pi_{\b a}^\perp +  \Pi_{\b b}^\perp)~,
\fe
where we defined $\delta_n(\b a,\b b) := \Vert  \Pi_{\b a}  \Pi_{\b b} -  \Pi_{\b a} \wedge  \Pi_{\b b} \Vert$.

Given $\b a$ and $\b b$ such that $\Vert \b a - \b b\Vert_{\infty} = 1$, let $k$ be the number of components/coordinates in which they differ (which can be $1$, $2$, or $3$). Then, we can write $\delta_n(\b a,\b b) = \delta_{n,k}$ because $\delta_n(\b a,\b b)$ depends only on $k$ (in addition to $n$) due to translation and rotational invariance. It follows that
\ie
\sum_{\b a,\b b: \Vert \b a - \b b\Vert_{\infty} = 1} \{  \Pi_{\b a}^\perp, \Pi_{\b b}^\perp \} \ge -(6\delta_{n,1} + 12 \delta_{n,2} + 8 \delta_{n,3}) \sum_{\b a}  \Pi_{\b a}^\perp~,
\fe
where we used the fact that for each $\b a$, there are $6$ $\b b$'s that differ from $\b a$ in one component, $12$ that differ in two components, and $8$ that differ in all three components. Defining $\delta_n := 6\delta_{n,1} + 12 \delta_{n,2} + 8 \delta_{n,3}$, we have
\ie
\bar H_L^2 &\ge (1-\delta_n) \sum_{\b a}  \Pi_{\b a}^\perp + \sum_{\b a,\b b:\Vert \b a - \b b\Vert_{\infty} > 1} \{  \Pi_{\b a}^\perp, \Pi_{\b b}^\perp \} \ge (1-\delta_n) \bar H_L~,
\fe
which implies
\ie
\bar \Delta_L \ge 1-\delta_n = 1-(6\delta_{n,1} + 12 \delta_{n,2} + 8 \delta_{n,3})~.
\fe
As we alluded to at the beginning of this subsection, we related the gap to the quantity $\delta_{n,k}$ associated with the ``local ground states'' in the big cubes.

In Appendix \ref{sec:3d-delta}, we define the quantity $\delta(A,B)$ \eqref{3d-delta-def} for any two overlapping rectangular boxes $A$ and $B$ in the lattice. In particular, when $A = C_{n\b a}$ and $B=C_{n\b b}$, we have $\delta_n(\b a,\b b) = \delta(C_{n\b a},C_{n\b b})$. In the appendix, we also derive the following upper bound on $\delta(A,B)$:
\ie\label{3d-upperbound-delta-maintext}
\delta(A,B) \le 4\cdot \frac{2^{V_\text{s}(A\cap B)/2}}{2^{V_\text{l}(A\cap B)/2}} + 3 \cdot \frac{2^{(V_\text{s}(A)+V_\text{s}(B))/2}}{2^{(V_\text{l}(A)+V_\text{l}(B))/2}}~,
\fe
where $V_\text{s}(A)$ and $V_\text{l}(A)$ denote the numbers of sites and links in the region $A$. Applying this bound to our case, where $A = C_{n\b a}$ and $B=C_{n\b b}$ with $\Vert \b a - \b b\Vert_\infty = 1$, we get
\ie
&\delta_{n,1} \le 4\cdot \frac{2^{n(2n)^2/2}}{2^{(3n-2)(2n)^2/2}} + 3 \cdot \frac{2^{(2n)^3}}{2^{3(2n-1)(2n)^2}}~,
\\
&\delta_{n,2} \le 4\cdot \frac{2^{n^2(2n)/2}}{2^{(6n-5)n^2/2}} + 3 \cdot \frac{2^{(2n)^3}}{2^{3(2n-1)(2n)^2}}~,
\\
&\delta_{n,3} \le 4\cdot \frac{2^{n^3/2}}{2^{3(n-1)n^2/2}} + 3 \cdot \frac{2^{(2n)^3}}{2^{3(2n-1)(2n)^2}}~.
\fe
For $n = 3$, we have
\ie
\delta_n \le 0.003 \implies \bar \Delta_L \ge 0.997~,
\fe
so $\bar H_L$ is gapped in the thermodynamic limit.

\section{Discussion and generalization to $\mathbb Z_N$}\label{sec:discuss}
In this work, we proposed and analyzed a deformation of the 3+1d lattice $\mathbb Z_2$ gauge theory that preserves the non-invertible Wegner duality symmetry at the self-dual point. We showed that there is a frustration-free point along this deformation where there are nine exactly degenerate ground states (on a periodic cubic lattice), one of which is a trivial product state and the rest are the ground states of the 3+1d toric code. Lastly, we proved that the deformed model at this frustration-free point is gapped in the thermodynamic limit. (Incidentally, we also proved the gap in a similar deformation of 1+1d critical Ising model, which was suggested by the numerics of \cite{OBrien:2017wmx}.)

\begin{figure}
\centering
\hfill \raisebox{-0.5\height}{\includegraphics[scale=0.23]{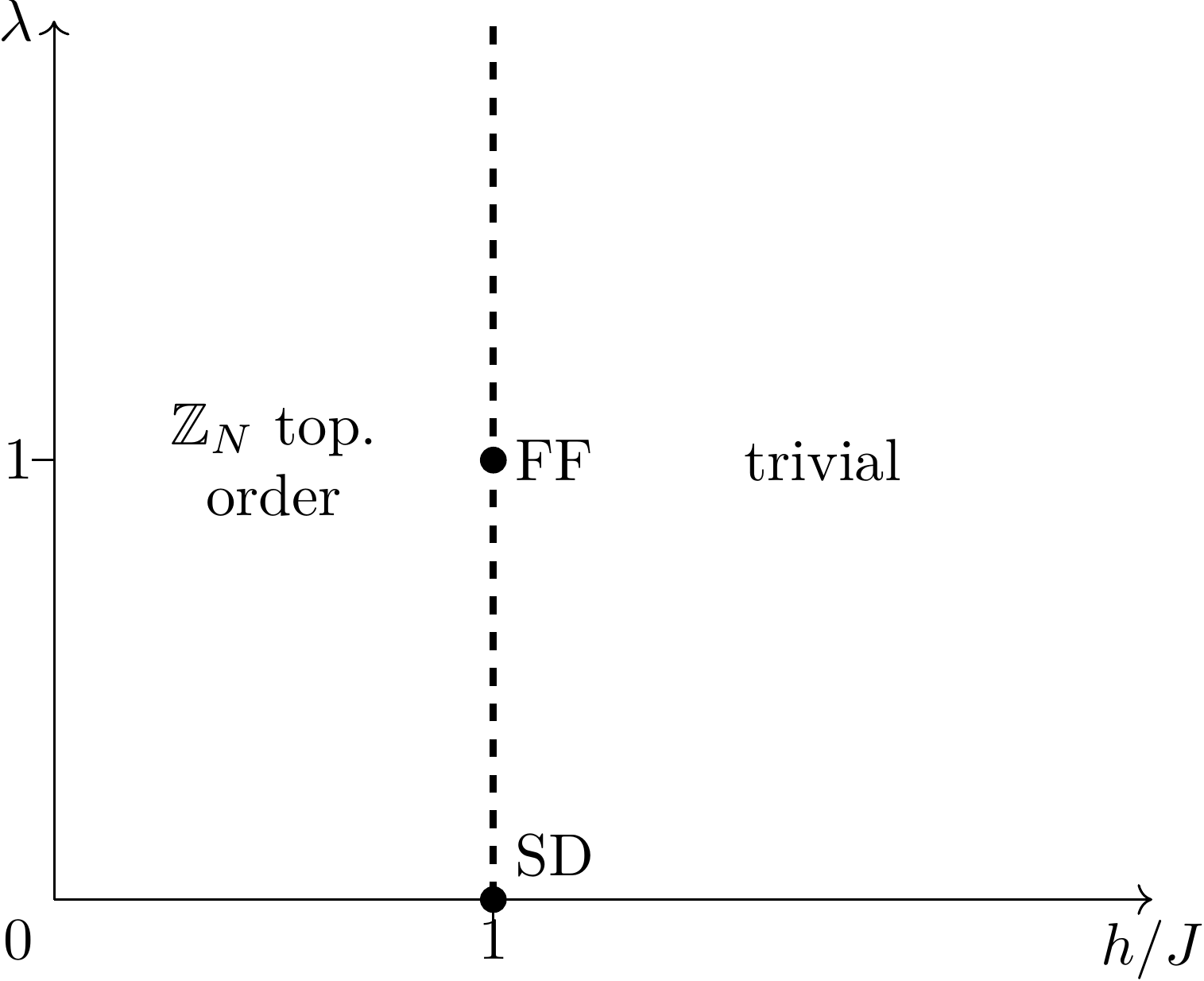}} \hfill \raisebox{-0.5\height}{\includegraphics[scale=0.23]{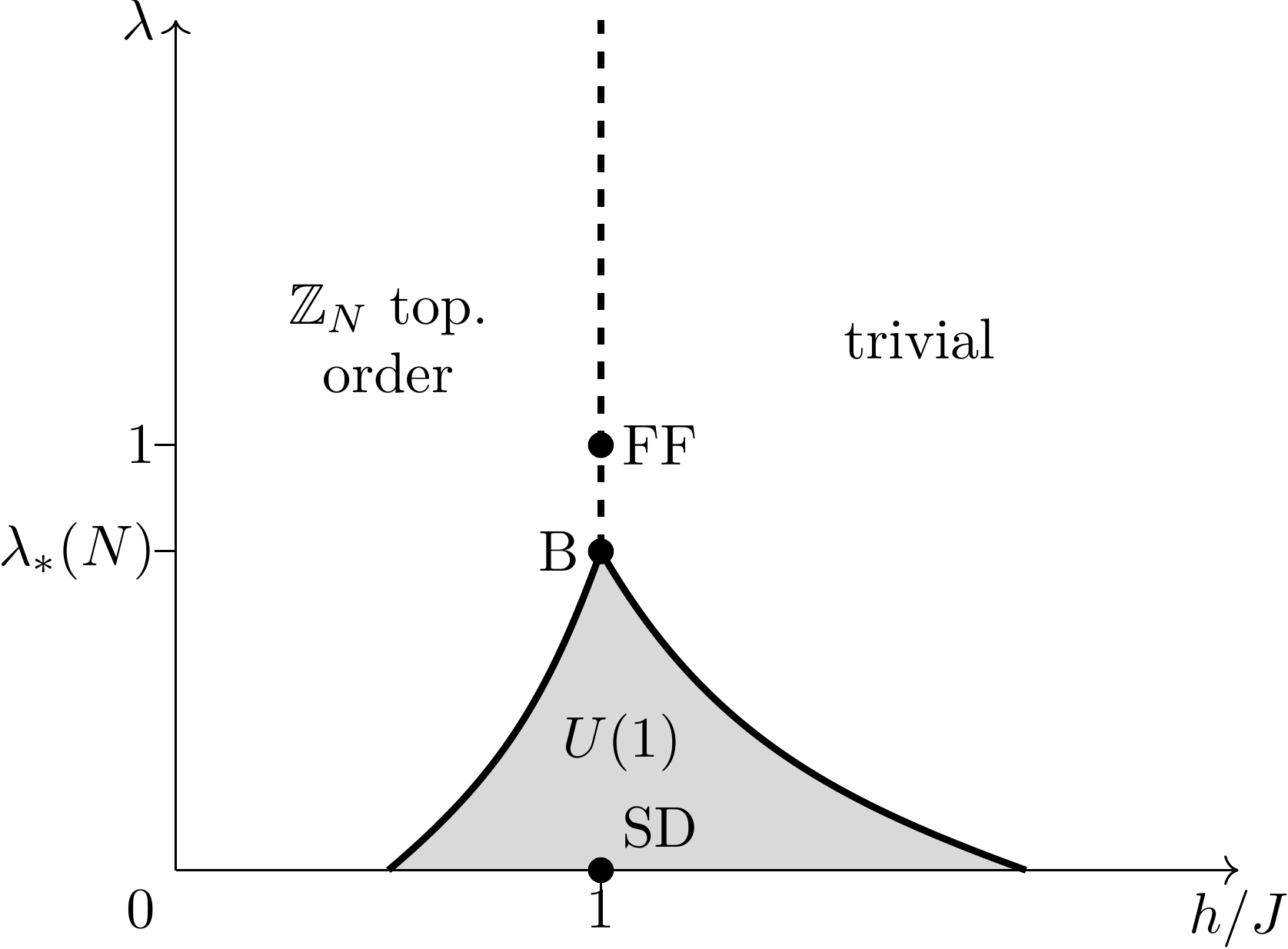}} \hfill ~
\\~\\
\hfill (a) ~~~~~~~~~~~\hfill~~~~~~~~~~~ (b) \hfill ~
\caption{Possible schematic phase diagram of the 3+1d lattice $\mathbb Z_N$ gauge theory with the Wegner duality-preserving deformation for (a) $N\le4$ and (b) $N>4$. Here, the unshaded (shaded) region represents a gapped (gapless) phase and the dashed (solid) line represents a first-order/gapped (second-order/higher-order/gapless) transition. The self-dual line $J=h$ has the non-invertible Wegner duality symmetry. There are two special points on this line: SD at $\lambda = 0$ is the self-dual point of the lattice $\mathbb Z_N$ gauge theory and FF at $\lambda = 1$ is the frustration-free point, which is the main focus of this paper. For $N>4$, there is another special point: B at $\lambda = \lambda_*(N)$ is the bifurcation point where the gapless window closes.\label{fig:phase-diag}}
\end{figure}

An important future direction is to understand the phase diagram of the deformed lattice $\mathbb Z_2$ gauge theory. Since both the self-dual point ($\lambda = 0$) and the frustration-free point ($\lambda = 1$) realize a gapped phase with spontaneously broken Wegner duality symmetry, it is likely that this phase persists for all $\lambda \ge 0$. So, we expect the phase diagram in the $\frac{h}{J}$-$\lambda$ plane to look like Figure \ref{fig:phase-diag}(a). This is unlike what happens in a similar deformation of the critical Ising model in 1+1d, where there is a transition from a gapless phase to a gapped phase along the self-dual line \cite{OBrien:2017wmx} (see Appendix \ref{sec:1d-deformation} for more details).

The phase diagram gets more interesting if we replace $\mathbb Z_2$ with $\mathbb Z_N$. It is not too hard to see that there is a similar deformation of the 3+1d lattice $\mathbb Z_N$ gauge theory that preserves the $\mathbb Z_N$ version of the non-invertible Wegner duality symmetry at the self-dual point. All our results generalize to this model straightforwardly. In particular, there is a frustration-free point ($\lambda = 1$) along the self-dual line $J=h$, where the deformed model is gapped and has $N^3+1$ exactly degenerate ground states (on a periodic cubic lattice), one of which is a trivial product state and the rest are the ground states of the 3+1d $\mathbb Z_N$ toric code. For $N\le 4$, since there is a first-order order transition between the two gapped phases at the self-dual point $(\lambda = 0)$ \cite{PhysRevD.20.1915}, we expect the phase diagram of the deformed lattice $\mathbb Z_N$ gauge theory to still look like Figure \ref{fig:phase-diag}(a).

On the other hand, for $N>4$, there is a gapless window between the trivial and topologically-ordered gapped phases of the lattice $\mathbb Z_N$ gauge theory \cite{PhysRevD.20.1915}. In particular, the self-dual point is within this gapless phase, so there must be a phase transition along self-dual line $J=h$ from the self-dual point ($\lambda = 0$) to the frustration-free point ($\lambda = 1$). This suggests that, for $N>4$, the gapless window shrinks along the deformation, and closes at a ``bifurcation point'' on the self-dual line $J=h$ at some $\lambda = \lambda_*(N)$ between $0$ and $1$. Since the gapless window at $\lambda = 0$ is described by the 3+1d pure $U(1)$ gauge theory (Maxwell theory) \cite{PhysRevD.20.1915}, it is reasonable to assume that the entire gapless region is captured by it. Therefore, we conjecture that the phase diagram in the $\frac{h}{J}$-$\lambda$ plane looks like Figure \ref{fig:phase-diag}(b).

It would be interesting to understand the field theory at the bifurcation point. The left and right solid lines in Figure \ref{fig:phase-diag}(b) are Higgs transitions caused by electrically and magnetically charged matter with electric charge $N$ and magnetic charge $1$, respectively. So, naively, one might require both electrically and magnetically charged matter to capture the bifurcation point.\footnote{We thank Kantaro Ohmori for discussions on this point.}

It is also worth exploring the scope of the techniques we developed to prove the exact nine-fold degeneracy at the frustration-free point of the deformed lattice $\mathbb Z_2$ gauge theory. Our arguments are entirely combinatorial, with connections to homology, suggesting that they should be applicable to a much broader class of frustration-free models. We leave these questions for future investigations.

\section*{Acknowledgements}

We are grateful to Abhinav Prem, Shu-Heng Shao, and Nathanan Tantivasadakarn for stimulating discussions. We thank Paul Fendley and Kantaro Ohmori for comments on the draft. PG thanks Shu-Heng Shao and Nathanan Tantivasadakarn for collaboration on a related project \cite{Gorantla:2024ocs}. PG and TCH were supported by the Simons Collaboration on Global Categorical Symmetries. The authors of this paper were ordered alphabetically.

\appendix

\section{Exact ground states without Gauss law}\label{app:deformed-nogausslaw}

In this appendix, we construct the exact ground states of the Hamiltonian $H''$ \eqref{3d-deformedH''}. In this case, the only constraints on the ground states are $P_\ell Q_p = 0$ for all $\ell \perp p$, i.e., the ground states do not have to obey the Gauss law. As expected, there are a lot more ground states; they are given by
\ie\label{groundstates-nogausslaw}
&|{+}\cdots{+}\>~,
\\
&|\mathtt S;\zeta\> := 2\cdot 2^{V/2} \prod_{i<j} \left( \frac{1+(-1)^{\zeta_{ij}}\eta_{ij}}2 \right) \prod_s \left(\frac{1+(-1)^{\delta_{\mathtt S}(s)}G_s}{2}\right) |0\cdots0\>~,
\fe
where $\mathtt S$ is any subset of all the sites in the lattice, $\zeta = (\zeta_{xy},\zeta_{yz},\zeta_{zx})\in \{0,1\}^3$, and $\delta_{\mathtt S}(s)$ is the indicator function of $\mathtt S$ on the sites of the lattice, i.e., $\delta_{\mathtt S}(s) = 1$ when $s\in \mathtt S$ and $0$ when $s\notin \mathtt S$. The states $|\mathtt S;\zeta\>$ are the analogues of $|\zeta\>$ in \eqref{zeta-states} in the absence of Gauss law constraints. Note that the relation $\prod_s G_s = 1$ implies $|\mathtt S;\zeta\> = 0$ when $|\mathtt S|$ is odd, so the nontrivial states are those with $|\mathtt S|$ even. Therefore, the total number of ground states is
\ie\label{noof-gndstates-nogausslaw}
1+8 \sum_{n=0}^{\lfloor V/2 \rfloor} \binom{V}{2n} = 1 + 8 \cdot 2^{V-1} =1+2^{V+2}~.
\fe

The argument in proving that these are the only ground states is similar to the one in Section \ref{sec:3d-proofofgsd}, so we highlight only the differences here. First, the decomposition of the state $|\psi\>$ in \eqref{arbitrarystate'} is modified to
\ie
|\psi\> = \sum_{\sigma\in\{+,-\}^{3V}} \psi_\sigma|\sigma\>~,
\fe
where there is no prime on the summation, i.e., we include sign configurations $\sigma$ that need not obey the Gauss law. For any configuration $\sigma$, define the set $\mathtt S(\sigma):=\{ s: \prod_{\ell\ni s} \sigma_\ell = -\}$, i.e., $\mathtt S(\sigma)$ is the set of sites where the Gauss law is violated in $|\sigma\>$. Note that $|\mathtt S(\sigma)|$ is even for any $\sigma$ because $\prod_s G_s = 1$ on a periodic lattice.

Consider the pictorial representation where links with $-$ are colored blue and links with $+$ are uncolored. Since there is no Gauss law constraint, the number of blue links around a site need not be even. In fact, the sites in $\mathtt S(\sigma)$ are exactly those with odd number of blue links around them. So the blue links in any configuration $\sigma \ne {+}\cdots{+}$ form not only closed cycles but also open paths with end points in the set $\mathtt S(\sigma)$.

The moves \eqref{3d-moves}, \eqref{translate}, \eqref{rotate}, \eqref{absorb} preserve the set $\mathtt S(\sigma)$ and also the $\mathbb Z_2^{(1)}$ charges of $\sigma$. Therefore, using a homological argument similar to the one in Section \ref{sec:3d-proofofgsd}, for any $\sigma\ne {+}\cdots{+}$ and $\sigma'\ne {+}\cdots{+}$, we can show that $\psi_\sigma = \psi_{\sigma'}$ whenever $\mathtt S(\sigma) = \mathtt S(\sigma')$, and $\sigma$ and $\sigma'$ carry the same charges under the $\mathbb Z_2$ 1-form symmetry (i.e., whenever $G_s|\sigma\> = G_s |\sigma'\>$ for all $s$ and $\eta_{ij}|\sigma\> = \eta_{ij}|\sigma'\>$ for all $i<j$). It follows that any ground state of $H''$ \eqref{3d-deformedH''} is a linear combination of the states in \eqref{groundstates-nogausslaw}.

Analogous to the toric code ground states $|\xi\>$, we can transform to the basis:
\ie
|\mathtt S; \xi\> &:= \frac1{2\sqrt2} \sum_{\zeta\in\{0,1\}^3} (-1)^{\sum_{ij} \zeta_{ij} \xi_{ij}} |\mathtt S;\zeta\>
\\
&= 2^{(V-1)/2} \prod_{i<j} \eta_{ij}^{\xi_{ij}} \prod_s \left(\frac{1+(-1)^{\delta_{\mathtt S}(s)}G_s}{2}\right) |0\cdots0\>~,
\fe
where $\xi = (\xi_{xy},\xi_{yz},\xi_{zx}) \in \{0,1\}^3$. More drastically, we can transform this basis further to a product basis:\footnote{In showing that the second line of \eqref{prod-groundstates-nogausslaw} follows from the first line, it is useful to note that $|S\cap \mathtt S| = \sum_s \delta_S(s) \delta_{\mathtt S}(s)$.}
\ie\label{prod-groundstates-nogausslaw}
|S;\xi\> &:= \frac1{2^{(V-1)/2}} \sum_{\mathtt S:|\mathtt S|\text{ is even}} (-1)^{|S\cap \mathtt S|} |\mathtt S;\xi\>
\\
&=\prod_{i<j} \eta_{ij}^{\xi_{ij}} \prod_{s\in S} G_s |0\cdots0\>~,
\fe
where $S$ is any subset of all the sites in the lattice. Note that the relation $\prod_s G_s = 1$ implies $|S;\xi\> = |S^\complement;\xi\>$, where $S^\complement := \{s: s\notin S\}$. Therefore, the number of ground states is
\ie
1+ \frac12 \cdot 8 \cdot 2^V = 1+ 2^{V+2}~,
\fe
which matches with \eqref{noof-gndstates-nogausslaw}.

\section{Deformation of 1+1d critical Ising model}\label{app:deformed-Ising}
In this appendix, we consider a deformation of the 1+1d critical Ising model that preserves the $\mathbb Z_2$ symmetry and the non-invertible Kramers-Wannier (KW) duality symmetry. This deformation was proposed in \cite{OBrien:2017wmx}. 
See \cite{2015PhRvL.115p6401R,2015PhRvB..92w5123R,Seiberg:2024gek} for other deformations preserving both of these symmetries.

Recall that the Hamiltonian of the 1+1d transverse-field Ising model is
\ie\label{1d-H}
H = -J\sum_{i=1}^L Z_i Z_{i+1} - h \sum_{i=1}^L X_i~.
\fe
Here, $L$ is the length of the 1d chain and we assume periodic boundary conditions. Each site $i$ hosts a qubit $\mathcal H_i = \mathbb C^2$, so the total Hilbert space is the tensor product $\mathcal H = \bigotimes_i \mathcal H_i$.

For general $J,h$, the Hamiltonian is invariant under a $\mathbb Z_2$ symmetry generated by
\ie\label{1d-eta-def}
\eta = \prod_i X_i~.
\fe
It is also invariant under lattice translations. At the critical point $J=h$, it furthermore has a non-invertible KW duality symmetry generated by the operator $\mathsf D$ that acts on the terms of the Hamiltonian as
\ie\label{1d-D-action}
\mathsf D X_i = Z_i Z_{i+1} \mathsf D~,\qquad \mathsf D Z_i Z_{i+1} = X_{i+1} \mathsf D~.
\fe
For various explicit expressions of $\mathsf D$, and the algebra of all the symmetry operators, see \cite{Aasen:2016dop,Tantivasadakarn:2021vel,Li:2023ani,Chen:2023qst,Seiberg:2023cdc,Seiberg:2024gek,Gorantla:2024ocs}.

The transverse-field Ising model is exactly solvable using Jordan-Wigner transformation. It has two gapped phases: a spontaneously broken (ordered or ferromagnetic) phase for $J>h$ and a symmetry preserving (disordered or paramagnetic) phase for $J<h$. The transition between them is of second order ($c=1/2$ Ising CFT) and it occurs exactly at $J=h$ due to KW duality.

The disordered phase has a unique ground state---for instance, at $J=0$, the ground state is
\ie\label{+state}
|{+}\cdots{+}\>~.
\fe
On the other hand, the ordered phase has two ground states---in particular, at $h=0$, the two ground states are
\ie\label{01states}
|0\cdots0\>~,\qquad |1\cdots1\>~.
\fe
While $\eta$ permutes the ground states within a phase, the KW duality operator $\mathsf D$ exchanges the two phases, i.e.,
\ie\label{1d-actiononstates}
&\eta |0\cdots0\> = |1\cdots1\>~,\qquad \eta |1\cdots1\> = |0\cdots\>~,\qquad && \eta |{+}\cdots{+}\> = |{+}\cdots{+}\>~,
\\
&\mathsf D|0\cdots0\> = \mathsf D |1\cdots1\> = |{+}\cdots{+}\>~,\qquad && \mathsf D|{+}\cdots{+}\> = |0\cdots0\> + |1\cdots1\>~.
\fe
If these three states were ground states of a gapped Hamiltonian, then that model would realise a spontaneously broken KW duality symmetry as implied by \eqref{1d-actiononstates}. Such a Hamiltonian was constructed in \cite{OBrien:2017wmx}, and we will analyze it in the rest of this appendix.

\subsection{Kramers-Wannier duality preserving deformation}\label{sec:1d-deformation}
Consider the following deformation of the critical Ising model that preserves both the $\mathbb Z_2$ symmetry and the KW duality symmetry \cite{OBrien:2017wmx}:
\ie\label{1d-deformedHlambda}
H_\lambda = -J \left[\sum_i (Z_i Z_{i+1} + X_i) - \frac{\lambda}2 \sum_i (X_{i-1} Z_i Z_{i+1} + Z_{i-1} Z_i X_{i+1})\right]~,
\fe
The phase diagram of this model with parameter $\lambda$ was studied numerically in \cite{OBrien:2017wmx}. It is gapless for $\lambda < \lambda_c$, described by $c=1/2$ Ising CFT, and gapped with three ground states for $\lambda > \lambda_c$. The transition point $\lambda_c \approx 0.856$ is captured by the $c=7/10$ tricritical Ising CFT. All of these phases agree with the generalized LSM-type constraints for the non-invertible KW duality symmetry \cite{Levin:2019ifu,Seiberg:2024gek}.

A particularly interesting point on this phase diagram is $\lambda = 1$. At this point, this model has an exact three-fold degeneracy even at finite $L$ \cite{OBrien:2017wmx}. In fact, the three ground states are precisely the three product states in \eqref{+state} and \eqref{01states}. Moreover, it falls within the gapped phase mentioned in the last paragraph.

Below, we give an alternative proof of the exact three-fold degeneracy at $\lambda = 1$---which is conceptually easier to generalize to a similar model in 3+1d in Section \ref{sec:3d-deform}---and we give an analytic proof of the gap at $\lambda=1$ in the thermodynamic limit.

\subsection{Exact ground states}
First, as noticed by O'Brien and Fendley in \cite{OBrien:2017wmx}, it is convenient to write the Hamiltonian at $\lambda=1$ as
\ie\label{1d-deformedH}
H' &= \frac{J}{2} \sum_i \left[ (1-X_{i-1}) (1-Z_i Z_{i+1}) + (1-Z_{i-1} Z_i) (1-X_{i+1}) \right]
\\
&= 2J \sum_i \left( P_{i-1} Q_{i,i+1} + Q_{i-1,i} P_{i+1}\right)~,
\fe
which differs from $H_{\lambda=1}$ only by a scalar multiple of the identity operator. Here, $P_i := \frac12(1-X_i)$ and $Q_{i,i+1} := \frac12(1-Z_i Z_{i+1})$ are orthogonal projection operators. Moreover, $P_{i-1}$ and $P_{i+2}$ commute with $Q_{i,i+1}$. Therefore, each term in the Hamiltonian $H'$ is an orthogonal projection operator. In particular, each term is positive semi-definite, so $H'$ is also positive semi-definite, and hence its energies are all nonnegative.

It is easy to see that
\ie\label{1d-groundstates}
H' |0\cdots0\> = 0~,\qquad H' |1\cdots1\> = 0~,\qquad H' |{+}\cdots{+}\> = 0~.
\fe
So the ground state energy is $0$ and there are at least three ground states. In particular, this means that $H'$ is frustration-free.

As we discussed above, $|0\cdots0\>$ and $|1\cdots1\>$ are the ground states of the Ising model \eqref{1d-H} at $h=0$ and $|{+}\cdots{+}\>$ is the ground state at $J=0$. What we have here is a gapped, frustation-free model $H'$ that preserves the $\mathbb Z_2$ symmetry and the KW duality symmetry, and has these three states as its exact ground states even at finite $L$.

\subsection{Proof of three-fold degeneracy}\label{sec:anotherproofGSD1d}
Here, we show that the only ground states of $H'$ \eqref{1d-deformedH} are the ones in \eqref{1d-groundstates}. Our proof is different from the one in \cite{OBrien:2017wmx}, and it generalizes easily to the Wegner duality-preserving deformation in 3+1d in Section \ref{sec:3d-deform}.

Any state $|\psi\>$ that is annihilated by $H'$, i.e., a ground state, must be annihilated by each term in the Hamiltonian due to positive semi-definiteness.\footnote{If $A$ and $B$ are Hermitian positive semi-definite matrices, then $(A+B) |v\> = 0\implies \< v| A |v\> + \< v| B |v\> = 0 \implies A |v\> = 0$ and $B |v\> = 0$.} That is,
\ie\label{eachtermzero}
P_{i-1} Q_{i,i+1} |\psi\> = 0~, \qquad Q_{i-1,i} P_{i+1} |\psi\> = 0~.
\fe
for all $i$.

Let us decompose $|\psi\>$ in the eigenbasis of $X_i$'s:
\ie\label{decomp}
|\psi\> = \sum_{\sigma \in \{+,-\}^L} \psi_\sigma |\sigma\>~,
\fe
where $\sigma$ is a sign configuration on the sites and $\psi_\sigma \in \mathbb C$ is the ``weight'' of the sign configuration $\sigma$. Consider the action of the operator $P_{i-1} Q_{i,i+1}$ on the basis state $|\sigma\>$,
\ie\label{PQonbasis}
&P_{i-1} Q_{i,i+1} |\cdots\sigma_i\sigma_{i+1}\cdots\>
\\
&= \begin{cases}
0~,&\sigma_{i-1} = +~,
\\
\frac12(|\cdots\sigma_i\sigma_{i+1}\cdots\> - |\cdots\bar\sigma_i\bar\sigma_{i+1}\cdots\>)~,&\sigma_{i-1} = -~,
\end{cases}
\fe
where $\bar\sigma_i = -\sigma_i$. So, the constraint $P_{i-1} Q_{i,i+1} |\psi\> = 0$ gives the relation
\ie
\psi_{\cdots\sigma_i\sigma_{i+1}\cdots} = \psi_{\cdots\bar\sigma_i\bar\sigma_{i+1}\cdots}~,\quad \text{if}\quad \sigma_{i-1} = -~.
\fe
Similarly, the constraint $Q_{i-1,i} P_{i+1} |\psi\> = 0$ gives the relation
\ie
\psi_{\cdots\sigma_{i-1}\sigma_i\cdots} = \psi_{\cdots\bar\sigma_{i-1}\bar\sigma_i\cdots}~,\quad \text{if}\quad \sigma_{i+1} = -~.
\fe
Note that all relations are among $\psi_\sigma$'s with $\sigma \ne {+}\cdots{+}$, i.e., there is no relation involving $\psi_{{+}\cdots{+}}$.

These relations can be restated as follows: given any $\sigma \ne {+}\cdots{+}$, let $\sigma'$ be the configuration obtained by flipping the two signs to the left or right (in a periodic way) of any $-$ in $\sigma$. Then, the above constraints imply that $\psi_{\sigma'} = \psi_\sigma$ for any such $\sigma'$. In fact, $\psi_{\sigma'} = \psi_\sigma$ whenever $\sigma'$ can be obtained from $\sigma$ by not just one flip but a sequence of flips.

We claim that any $\sigma' \ne {+}\cdots{+}$ can be obtained from any $\sigma \ne {+}\cdots{+}$ by a sequence of flips if and only if the numbers of $-$'s in $\sigma$ and $\sigma'$ have the same parity, i.e., if and only if $\eta|\sigma\> = \eta|\sigma'\>$. That this is necessary is clear because each flip flips two signs, so the parity of number of $-$'s is preserved under a flip. To show that this is sufficient, it is enough to show that every $\sigma\ne {+}\cdots{+}$ with odd (resp. even) number of $-$'s can be reduced to ${-}{+}\cdots{+}$ (resp. ${-}{-}{+}\cdots{+}$) by a sequence of flips.

Before proving the last statement, let us establish some rules.
\begin{itemize}
\item \underline{Rule 1}: a pair of opposite signs on sites $i$ and $i+1$ can always be flipped without changing any other signs:
\ie
\text{i}:\quad & {-}{\underset{i}{+}}{-}{-} \quad\overset{Q_{i,i+1}P_{i+2}}\longleftrightarrow\quad {-}{\underset{i}{-}}{+}{-}~,
\\
\text{ii}:\quad & {+}{\underset{i}{+}}{-}{-} \quad\overset{Q_{i,i+1}P_{i+2}}\longleftrightarrow\quad {+}{\underset{i}{-}}{+}{-}~,
\\
\text{iii}:\quad & {-}{\underset{i}{+}}{-}{+} \quad\overset{P_{i-1}Q_{i,i+1}}\longleftrightarrow\quad {-}{\underset{i}{-}}{+}{+}~,
\\
\text{iv}:\quad & {+}{\underset{i}{+}}{-}{+} \quad\overset{Q_{i-1,i}P_{i+1}}\longleftrightarrow\quad {-}{\underset{i}{-}}{-}{+} \quad\overset{P_iQ_{i+1,i+2}}\longleftrightarrow\quad {-}{\underset{i}{-}}{+}{-}
\\
& \qquad\overset{Q_{i,i+1}P_{i+2}}\longleftrightarrow\quad {-}{\underset{i}{+}}{-}{-} \quad\overset{Q_{i-1,i}P_{i+1}}\longleftrightarrow\quad {+}{\underset{i}{-}}{-}{-} \quad\overset{P_iQ_{i+1,i+2}}\longleftrightarrow\quad {+}{\underset{i}{-}}{+}{+}~,
\fe
where we show only the sites $i-1$, $i$, $i+1$, and $i+2$. (While i, ii, and iii involve only one flip, iv involves a sequence of flips because the signs adjacent to $i$ and $i+1$ are both $+$.) It is useful to interpret the above flips as follows: a $-$ can always ``move'' to the left or right as long as it does not encounter another $-$.
\item \underline{Rule 2}: a contiguous group of three or more $-$'s can be reduced by two $-$'s at a time; for example,
\ie
{-}{\underset{i}{-}}{-}{+} \quad\overset{P_{i-1}Q_{i,i+1}}\longleftrightarrow\quad {-}{\underset{i}{+}}{+}{+}~.
\fe
\end{itemize}
Now, given a $\sigma\ne {+}\cdots{+}$, we use Rule 1 to ``move'' all the $-$'s to a single contiguous group on the left. Then, we use Rule 2 to reduce it to ${-}{+}\cdots{+}$ or ${-}{-}{+}\cdots{+}$ depending on the parity of the number of $-$'s in $\sigma$. This proves the claim.

It follows that, for any $\sigma \ne {+}\cdots{+}$, $\psi_{\sigma} = \psi_{{-}{-}{+}\cdots{+}}$ if $\eta|\sigma\> = |\sigma\>$ and $\psi_{\sigma} = \psi_{{-}{+}\cdots{+}}$ if $\eta|\sigma\> = -|\sigma\>$. Hence, any state $|\psi\>$ that satisfies \eqref{eachtermzero} takes the form
\ie
|\psi\> = \alpha |{+}\cdots{+}\> + \psi_{{-}{-}{+}\cdots{+}} \sum_{\sigma\in\{+,-\}^L \atop \eta|\sigma\> = |\sigma\>} |\sigma\> + \psi_{{-}{+}\cdots{+}} \sum_{\sigma\in\{+,-\}^L \atop \eta|\sigma\> = - |\sigma\>} |\sigma\>~,
\fe
where $\alpha := \psi_{{+}\cdots{+}} - \psi_{{-}{-}{+}\cdots{+}}$. It is easy to show that
\ie
&\sum_{\sigma\in\{+,-\}^L \atop \eta|\sigma\> = |\sigma\>} |\sigma\> = \frac{2^{L-1}}{2^{L/2}} (|0\cdots0\> + |1\cdots1\>)~,
\\
&\sum_{\sigma\in\{+,-\}^L \atop \eta|\sigma\> = -|\sigma\>} |\sigma\> = \frac{2^{L-1}}{2^{L/2}} (|0\cdots0\> - |1\cdots1\>)~,
\fe
so we can write $|\psi\>$ equivalently as 
\ie
|\psi\> = \alpha|{+}\cdots{+}\> + \beta|0\cdots0\> + \gamma|1\cdots1\>~,
\fe
for some $\beta,\gamma\in \mathbb C$. This shows that the three states in \eqref{1d-groundstates} are the only ground states of $H'$.

\subsection{Proof of gap}\label{sec:1d-proofofgap-martingale}
In this appendix, we prove that the Hamiltonian $H'$ \eqref{1d-deformedH} is gapped in the thermodynamic limit using the martingale method \cite{Fannes:1992,Nachtergaele:1996,Kastoryano_2018}. The existence of gap was suggested by the numerics in \cite{OBrien:2017wmx}, and here, we give an analytic proof.

For simplicity, we assume that $L$ is a multiple of an integer $n\ge2$, which we will fix later. We further assume that $L\ge 3n$. (The following proof can be generalized to $L$ that are not multiples of $n$.) We first coarse-grain the lattice by a factor of $n$ so that each new site, labelled by $I = 1,\ldots, L/n$, contains the $n$ sites $i=n(I-1)+1,\ldots,nI$. In other words, each new site $I$ is an interval of $n$ consecutive sites with a local Hilbert space $\mathcal H_I := \bigotimes_{i=n(I-1)+1}^{nI} \mathcal H_i$ of dimension $2^n$. Consider the local Hamiltonian
\ie
h_{I,I+1} := \sum_{i=n(I-1)+1}^{n(I+1)-2} (P_i Q_{i+1,i+2} + Q_{i,i+1} P_{i+2})~,
\fe
which acts nontrivially on the $2n$ qubits in $\mathcal H_I \otimes \mathcal H_{I+1}$. Since $n\ge 2$, using the arguments in Appendix \ref{sec:anotherproofGSD1d}, it can be shown that each $h_{I,I+1}$ has exactly three ground states given by
\ie
|{+}\cdots{+}\>_I |{+}\cdots{+}\>_{I+1}~,\qquad |0\cdots0\>_I |0\cdots0\>_{I+1}~,\qquad |1\cdots1\>_I |1\cdots1\>_{I+1}~,
\fe
where the subscript $I$ means that the state belongs to $\mathcal H_I$.

Each term in $H'$ appears in at least one and at most two of the $h_{I,I+1}$'s. So, we have
\ie
H'_L \ge \frac12 \cdot 2J \sum_{I=1}^{L/n} h_{I,I+1}~,
\fe
where we introduced the subscript $L$ on $H'$ for clarity. Let $ \Pi_{I,I+1}$ be the projector onto the (nontrivial) kernel of $h_{I,I+1}$, and $ \Pi_{I,I+1}^\perp := 1 -  \Pi_{I,I+1}$. Then, $h_{I,I+1} \ge \epsilon_n  \Pi_{I,I+1}^\perp$, where $\epsilon_n>0$ is the smallest positive eigenvalue of $h_{I,I+1}$, which is independent of $I$ due to translation invariance, but can depend on $n$, the number of qubits in each new site $I$. It follows that
\ie
H'_L \ge J \epsilon_n \bar H_L~,\qquad \text{where} \qquad \bar H_L := \sum_{I=1}^{L/n}  \Pi_{I,I+1}^\perp~.
\fe
Frustration-freeness implies that $H'_L$ has the same ground states as $\bar H_L$, so it suffices to show that $\bar H_L$ is gapped, as long as $n$ is finite (so that $\epsilon_n > 0$).

Following Knabe's argument \cite{Knabe:1988}, our goal is to show that there is a constant $\Delta_0>0$ independent of $L$ such that $\bar H_L^2 \ge \Delta_0 \bar H_L$ for all $L>L_0$ that are multiples of $n$ for some $L_0$. Since $\bar H_L$ is positive semi-definite, it follows that $\bar \Delta_L \ge \Delta_0 > 0$ for all such $L$, thereby showing that $\bar H_L$ is gapped.

Consider the square of the Hamiltonian,
\ie
\bar H_L^2 = \sum_I \Pi_{I,I+1}^\perp + \sum_I \{ \Pi_{I,I+1}^\perp, \Pi_{I+1,I+2}^\perp\} + \sum_{I,I':|I-I'|>1}  \Pi_{I,I+1}^\perp \Pi_{I',I'+1}^\perp~.
\fe
The first term is simply $\bar H_L$. Since $ \Pi_{I,I+1}^\perp$ commutes with $ \Pi_{I',I'+1}^\perp$ if and only if $|I-I'|\ne1$, the last term is positive semi-definite, but the second term is not necessarily positive semi-definite.

We take care of the second term using the martingale method, which relies on the following mathematical facts \cite[Lemma 6.3]{Fannes:1992}: any two orthogonal projections $ \Pi_1$ and $ \Pi_2$ satisfy
\ie
&\{ \Pi_1, \Pi_2\} \ge - \Vert  \Pi_1  \Pi_2 -  \Pi_1 \wedge  \Pi_2 \Vert~( \Pi_1 +  \Pi_2)~,
\fe
and
\ie
&\Vert  \Pi_1  \Pi_2 -  \Pi_1 \wedge  \Pi_2 \Vert = \Vert  \Pi_1^\perp  \Pi_2^\perp -  \Pi_1^\perp \wedge  \Pi_2^\perp \Vert~,
\fe
where $\Vert \cdot \Vert$ denotes the operator norm, $ \Pi_1 \wedge  \Pi_2$ is the orthogonal projection onto $\im( \Pi_1) \cap \im( \Pi_2)$, and $ \Pi_{1,2}^\perp := 1- \Pi_{1,2}$. In our case, with $ \Pi_1 =  \Pi_{I,I+1}^\perp$ and $ \Pi_2 =  \Pi_{I+1,I+2}^\perp$, the inequality is
\ie
\{ \Pi_{I,I+1}^\perp, \Pi_{I+1,I+2}^\perp\} \ge - \delta_n ( \Pi_{I,I+1}^\perp +  \Pi_{I+1,I+2}^\perp)~,
\fe
where we defined $\delta_n := \Vert  \Pi_{I,I+1}  \Pi_{I+1,I+2} -  \Pi_{I,I+1} \wedge  \Pi_{I+1,I+2} \Vert$, which is independent of $I$ due to translation invariance, but can depend on $n$. It follows that
\ie
\sum_I \{ \Pi_{I,I+1}^\perp, \Pi_{I+1,I+2}^\perp\} \ge -2\delta_n \sum_I  \Pi_{I,I+1}^\perp = -2\delta_n \bar H_L~,
\fe
and hence,
\ie
\bar H_L^2 \ge (1-2\delta_n) \bar H_L \implies \bar \Delta_L \ge 1-2\delta_n~.
\fe
Thus, we have related the gap to the quantity $\delta_n$ associated with the local ground states in the new sites.

In Appendix \ref{sec:1d-delta}, we define the quantity $\delta(A,B)$ \eqref{1d-delta-def} for any two overlapping connected intervals $A$ and $B$ in the lattice with at least $4$ sites in each interval and at least $2$ sites in their intersection. We also derive an upper bound on $\delta(A,B)$ in terms of number of sites in $A$, $B$, and $A\cap B$:
\ie\label{1d-delta-upperbound-maintext}
\delta(A,B) &\le \frac{8}{2^{(|A|+|B|)/2}} + \frac{4}{2^{|A\cap B|/2}}~.
\fe
In our case, $A$ (resp. $B$) is an interval of size $2n$ associated with the new sites $I$ and $I+1$ (resp. $I+1$ and $I+2$), and their intersection $A\cap B$ is the interval of size $n$ associated with the new site $I+1$. Then, the upper bound \eqref{1d-delta-upperbound-maintext} gives
\ie
\delta_n \le \frac{8}{2^n} + \frac{4}{2^{n/2}}~.
\fe
For $n= 8$, we have
\ie
\delta_n \le 0.3 \implies \bar \Delta_L \ge 0.4~,
\fe
so $\bar H_L$ is gapped in the thermodynamic limit.

\section{Upper bound on $\delta(A,B)$}\label{app:up-bound-delta}

In this appendix, we define and upper-bound the quantity $\delta(A,B)$ for the Hamiltonians $H'$ \eqref{1d-deformedH} in 1+1d and $H''$ \eqref{3d-deformedH''} in 3+1d, where $A$ and $B$ are two overlapping regions in the lattice.

\subsection{$\delta(A,B)$}\label{app:delta}
Given a frustration-free Hamiltonian $H$ on the lattice, one can define the ``local Hamiltonian'' $H_\Lambda$ on a finite local region $\Lambda$ as the sum of all the local terms in the Hamiltonian $H$ that are contained within $\Lambda$. It acts on the ``local Hilbert space'' $\mathcal H_\Lambda$. It is easy to see that $H_\Lambda$ is also frustration-free. The ``local ground states'' in the region $\Lambda$ are the ground states of $H_\Lambda$, i.e., those states that are annihilated by the terms of the Hamiltonian $H$ contained within $\Lambda$. Let $\Pi_\Lambda$ denote the orthogonal projection onto the ``local ground state space'' $\mathcal H_\Lambda^0$ in the region $\Lambda$.

For any two overlapping regions $A$ and $B$, the quantity $\delta(A,B)$ is defined as
\ie
\delta(A,B) := \Vert \Pi_A \Pi_B - \Pi_A \wedge \Pi_B \Vert~.
\fe
where $\Vert \cdot \Vert$ denotes the operator norm and $\Pi_A \wedge \Pi_B$ is the orthogonal projection onto $\mathcal H_A^0 \cap \mathcal H_B^0$. While frustration-freeness implies that $\mathcal H_A^0 \cap \mathcal H_B^0$ contains $\mathcal H_{A\cup B}^0$, it does not necessarily imply that they are equal. In some cases, including the Hamiltonians we are interested in, it turns out that $\mathcal H_A^0 \cap \mathcal H_B^0 = \mathcal H_{A\cup B}^0$, so the above quantity can also be written as
\ie
\delta(A,B) = \Vert \Pi_A \Pi_B - \Pi_{A \cup B} \Vert~.
\fe

By definition, $\delta(A,B)$ is determined entirely by the local ground states of the Hamiltonian in the regions $A$ and $B$. More importantly, it is intimately related to the gap of the Hamiltonian in the thermodynamic limit: roughly, there is a nonzero gap if and only if $\delta(A,B)$ decays with increasing size of the overlap $A\cap B$ for all sufficiently large regions $A$ and $B$. (For a more precise statement, see \cite{Kastoryano_2018}.) As we show below, this is precisely the case for the Hamiltonians $H'$ \eqref{1d-deformedH} in 1+1d and $H''$ \eqref{3d-deformedH''} in 3+1d.

\subsection{1+1d}\label{sec:1d-delta}
We now prove the upper bound \eqref{1d-delta-upperbound-maintext} on $\delta(A,B)$ for the Hamiltonian $H'$ \eqref{1d-deformedH} in 1+1d.

Let $\Lambda$ be a finite, connected interval in the lattice,\footnote{Here, $\Lambda$ is a \emph{connected interval} on the lattice if it satisfies the following: if two sites $i,i'$, with $i<i'$, belong to $\Lambda$, then all the sites $j$ in between them, i.e., $i<j<i'$, also belong to $\Lambda$.} and $\mathcal H_\Lambda := \bigotimes_{i\in\Lambda} \mathcal H_i$ be the tensor product Hilbert space on $\Lambda$. We use $|\Lambda|$ to denote the number of sites in $\Lambda$. Consider the restriction of the Hamiltonian \eqref{1d-deformedH} (up to factors of $2J$) to $\Lambda$,
\ie
h_\Lambda := \sum_{i\in\Lambda : i+2\in\Lambda} (P_i Q_{i+1,i+2} + Q_{i,i+1} P_{i+2})~.
\fe
Using an argument similar to the one in Appendix \ref{sec:anotherproofGSD1d}, we can show that, if $|\Lambda| \ge 4$, then $h_\Lambda$ has three ground states,
\ie
|\Plus\>_\Lambda := |{+}\cdots{+}\>_\Lambda~,\qquad |\Zero\>_\Lambda := |0\cdots0\>_\Lambda~,\qquad |\One\>_\Lambda := |1\cdots1\>_\Lambda~.
\fe
While $|\Zero\>_\Lambda$ and $|\One\>_\Lambda$ are orthogonal to each other, they are not orthogonal to $|\Plus\>_\Lambda$:
\ie
{}_\Lambda\<\Plus|\Zero\>_\Lambda = {}_\Lambda\<\Plus|\One\>_\Lambda = \frac1{2^{|\Lambda|/2}}~.
\fe
We use $\mathcal H^0_\Lambda$ to denote the subspace spanned by the three states above, and $ \Pi_\Lambda$ to denote the orthogonal projection onto $\mathcal H^0_\Lambda$, i.e., $\im( \Pi_\Lambda) = \mathcal H^0_\Lambda$.

Consider two overlapping connected intervals $A$ and $B$ such that $|A|\ge 4$, $|B|\ge 4$, and $|A\cap B| \ge 2$. Let $A' := A\smallsetminus B$ and $B' := B\smallsetminus A$. Note that $A'$, $A\cap B$, and $B'$ are also connected intervals. Then, the Hilbert space on $A\cup B$ factorizes as
\ie
\mathcal H_{A\cup B} = \mathcal H_{A'} \otimes \mathcal H_{A\cap B} \otimes \mathcal H_{B'}~,
\fe
and we have
\ie\label{1d-innerproduct}
&{}_A\<\Plus|\Plus\>_B = |\Plus\>_{B'}~{}_{A'}\<\Plus|~,
\\
&{}_A\<\mmm|\Plus\>_B = \frac1{2^{|A\cap B|/2}} |\Plus\>_{B'}~{}_{A'}\<\mmm|~,
\\
&{}_A\<\Plus|\mmm'\>_B = \frac1{2^{|A\cap B|/2}} |\mmm'\>_{B'}~{}_{A'}\<\Plus|~,
\\
&{}_A\<\mmm|\mmm'\>_B = \delta_{m,m'} |\mmm'\>_{B'}~{}_{A'}\<\mmm|~,
\fe
where we defined $|\mmm\> := |m\cdots m\>$ for $m=0,1$.

Note that $ \Pi_A  \Pi_{A\cup B} =  \Pi_{A\cup B}  \Pi_A =  \Pi_{A\cup B}$, and similarly for $ \Pi_B$, because of frustration-freeness. It follows that $ \Pi_A -  \Pi_{A\cup B}$ and $ \Pi_B -  \Pi_{A\cup B}$ are orthogonal projections too. Frustration-freeness also implies that $ \Pi_A \wedge  \Pi_B =  \Pi_{A\cup B}$.\footnote{It is obvious that $\mathcal H^0_{A\cup B} \subseteq \mathcal H^0_A \cap \mathcal H^0_B$ due to frustration-freeness. For the other direction, consider any $|\psi\> \in \mathcal H^0_A \cap \mathcal H^0_B$. Then, $ \Pi_A^\perp |\psi\>=0 \implies P_i Q_{i+1,i+2} |\psi\> = 0$ and $Q_{i,i+1} P_{i+2} |\psi\> = 0$ for all $i\in A$ such that $i+2\in A$ due to frustration-freeness, and similarly for all $i\in B$ such that $i+2\in B$. Since $|A\cap B| \ge 2$, we have $P_i Q_{i+1,i+2} |\psi\> = 0$ and $Q_{i,i+1} P_{i+2} |\psi\> = 0$ for all $i\in A\cup B$ such that $i+2\in A\cup B$, which means $ \Pi_{A\cup B}^\perp|\psi\> = 0$. Therefore, $\mathcal H^0_{A\cup B} = \mathcal H^0_A \cap \mathcal H^0_B$, or equivalently, $ \Pi_A \wedge  \Pi_B =  \Pi_{A\cup B}$.}

Consider the quantity
\ie\label{1d-delta-def}
\delta(A,B) := \Vert  \Pi_A  \Pi_B -  \Pi_A \wedge  \Pi_B \Vert = \Vert  \Pi_A  \Pi_B -  \Pi_{A\cup B} \Vert = \Vert ( \Pi_A - \Pi_{A\cup B})( \Pi_B -  \Pi_{A\cup B}) \Vert~,
\fe
Since the norm of product of orthogonal projections is always at most $1$, we have $0\le \delta(A,B) \le 1$. Below, we will show that
\ie\label{1d-delta-upper-bound}
\delta(A,B) &\le \frac{8}{2^{(|A|+|B|)/2}} + \frac{4}{2^{|A\cap B|/2}}~,
\fe
for all sufficiently large $A$ and $B$. In order to prove this bound, we consider an equivalent characterization of $\delta(A,B)$ in terms of states instead of projections:
\ie\label{1d-delta-state-def}
\delta(A,B) &= \sup\{ |\<\psi|\chi\>|: |\psi\>,|\chi\> \in \mathcal H_{A\cup B}\,,\, \<\psi|\psi\> \le 1\,,\,\<\chi|\chi\> \le 1\,,
\\
&\qquad \qquad \qquad \quad( \Pi_A - \Pi_{A\cup B})|\psi\> =|\psi\>\,,\, ( \Pi_B - \Pi_{A\cup B})|\chi\> =|\chi\>\}~.
\fe
Note that the condition $( \Pi_A - \Pi_{A\cup B})|\psi\> =|\psi\>$ means that $|\psi\> \in (\mathcal H^0_A \otimes \mathcal H_{B'}) \cap (\mathcal H^0_{A\cup B})^\perp$, i.e., $|\psi\>$ is a ground state in $A$ and it is orthogonal to the ground states in $A\cup B$,\footnote{Indeed, $ \Pi_A |\psi\> =  \Pi_A ( \Pi_A - \Pi_{A\cup B})|\psi\> = ( \Pi_A - \Pi_{A\cup B})|\psi\> = |\psi\>$, whereas $ \Pi_{A\cup B}|\psi\> =  \Pi_{A\cup B}( \Pi_A - \Pi_{A\cup B})|\psi\> = ( \Pi_{A\cup B} - \Pi_{A\cup B})|\psi\> = 0$.} and similarly for $|\chi\>$. So we write
\ie
&|\psi\> = |\Plus\>_A |\psi_+\>_{B'} + |\Zero\>_A |\psi_0\>_{B'} + |\One\>_A |\psi_1\>_{B'}~,
\\
&|\chi\> = |\chi_+\>_{A'} |\Plus\>_B + |\chi_0\>_{A'} |\Zero\>_B + |\chi_1\>_{A'} |\One\>_B~,
\fe
for some states $|\psi_{+,0,1}\>_{B'} \in \mathcal H_{B'}$ and $|\chi_{+,0,1}\>_{A'} \in \mathcal H_{A'}$, so that $|\psi\>$ and $|\chi\>$ are ground states in $A$ and $B$, respectively.

We now restrict them to be orthogonal to the ground states in $A\cup B$. First, we have
\ie\label{1d-ortho1}
&0 = {}_{A\cup B}\<\Plus|\psi\> = {}_{B'}\<\Plus|\psi_+\>_{B'} + \frac{1}{2^{|A|/2}} \sum_{m=0,1} {}_{B'}\<\Plus|\psi_m\>_{B'}~,
\\
&0 = {}_{A\cup B}\<\mmm|\psi\> = \frac{1}{2^{|A|/2}}~{}_{B'}\<\mmm|\psi_+\>_{B'} + {}_{B'}\<\mmm|\psi_m\>_{B'}~.
\fe
Similarly, we have
\ie\label{1d-ortho2}
&0 = {}_{A\cup B}\<\Plus|\chi\> = {}_{A'}\<\Plus|\chi_+\>_{A'} + \frac{1}{2^{|B|/2}} \sum_{m=0,1} {}_{A'}\<\Plus|\chi_m\>_{A'}~,
\\
&0 = {}_{A\cup B}\<\mmm|\chi\> = \frac{1}{2^{|B|/2}}~{}_{A'}\<\mmm|\chi_+\>_{A'} + {}_{A'}\<\mmm|\chi_m\>_{A'}~.
\fe
The constraints on the norms give 
\ie\label{1d-norm1}
1&\ge\<\psi|\psi\>
\\
&= {}_{B'}\<\psi_+|\psi_+\>_{B'} + \sum_{m=0,1} {}_{B'}\<\psi_m|\psi_m\>_{B'} + \frac{1}{2^{|A|/2}} \sum_{m=0,1} \left( {}_{B'}\<\psi_+|\psi_m\>_{B'} + {}_{B'}\<\psi_m|\psi_+\>_{B'} \right)
\\
&\ge \left( 1-\frac{2}{2^{|A|/2}} \right) {}_{B'}\<\psi_+|\psi_+\>_{B'} + \left( 1- \frac{1}{2^{|A|/2}} \right) \sum_{m=0,1}~{}_{B'}\<\psi_m|\psi_m\>_{B'}~,
\fe
and similarly,
\ie\label{1d-norm2}
1 \ge \left( 1-\frac{2}{2^{|B|/2}} \right) {}_{A'}\<\chi_+|\chi_+\>_{A'} + \left( 1- \frac{1}{2^{|B|/2}} \right) \sum_{m=0,1}~{}_{A'}\<\chi_m|\chi_m\>_{A'}~.
\fe

Now, the overlap between $|\psi\>$ and $|\chi\>$ is
\ie
\<\psi|\chi\> &= {}_{A'}\<\Plus|\chi_+\>_{A'}~{}_{B'}\<\psi_+|\Plus\>_{B'}
\\
&\quad + \frac{1}{2^{|A\cap B|/2}} \sum_{m=0,1} ~{}_{A'}\<\mathsf m|\chi_+\>_{A'}~{}_{B'}\<\psi_m|\Plus\>_{B'}
\\
&\quad + \frac{1}{2^{|A\cap B|/2}} \sum_{m=0,1} ~{}_{A'}\<\Plus|\chi_m\>_{A'}~{}_{B'}\<\psi_+|\mmm\>_{B'}
\\
&\quad + \sum_{m=0,1} ~{}_{A'}\<\mmm|\chi_m\>_{A'}~{}_{B'}\<\psi_m|\mmm\>_{B'}~,
\fe
where we used \eqref{1d-innerproduct}. Using \eqref{1d-ortho1} and \eqref{1d-ortho2}, we get
\ie
\<\psi|\chi\> &= \frac{1}{2^{(|A|+|B|)/2}} \sum_{m,m'=0,1}~{}_{A'}\<\Plus|\chi_{m'}\>_{A'}~{}_{B'}\<\psi_m|\Plus\>_{B'}
\\
&\quad + \frac{1}{2^{|A\cap B|/2}} \sum_{m=0,1} ~{}_{A'}\<\mathsf m|\chi_+\>_{A'}~{}_{B'}\<\psi_m|\Plus\>_{B'}
\\
&\quad + \frac{1}{2^{|A\cap B|/2}} \sum_{m=0,1} ~{}_{A'}\<\Plus|\chi_m\>_{A'}~{}_{B'}\<\psi_+|\mmm\>_{B'}
\\
&\quad + \frac{1}{2^{(|A|+|B|)/2}} \sum_{m=0,1} ~{}_{A'}\<\mmm|\chi_+\>_{A'}~{}_{B'}\<\psi_+|\mmm\>_{B'}~.
\fe
Taking the absolute value gives
\ie
|\<\psi|\chi\>| &\le \frac{1}{2^{(|A|+|B|)/2}} \sum_{m,m'=0,1}~|{}_{A'}\<\Plus|\chi_{m'}\>_{A'}|~|{}_{B'}\<\psi_m|\Plus\>_{B'}|
\\
&\quad + \frac{1}{2^{|A\cap B|/2}} \sum_{m=0,1} ~|{}_{A'}\<\mathsf m|\chi_+\>_{A'}|~|{}_{B'}\<\psi_m|\Plus\>_{B'}|
\\
&\quad + \frac{1}{2^{|A\cap B|/2}} \sum_{m=0,1} ~|{}_{A'}\<\Plus|\chi_m\>_{A'}|~|{}_{B'}\<\psi_+|\mmm\>_{B'}|
\\
&\quad + \frac{1}{2^{(|A|+|B|)/2}} \sum_{m=0,1} ~|{}_{A'}\<\mmm|\chi_+\>_{A'}|~|{}_{B'}\<\psi_+|\mmm\>_{B'}|~.
\fe
Let us take care of each line separately. First, we have
\ie
&\sum_{m,m'=0,1}~|{}_{A'}\<\Plus|\chi_{m'}\>_{A'}|~|{}_{B'}\<\psi_m|\Plus\>_{B'}|
\\
&\le \sqrt{2 \sum_{m'=0,1} |{}_{A'}\<\Plus|\chi_{m'}\>_{A'}|^2}~\sqrt{2 \sum_{m=0,1} |{}_{B'}\<\psi_m|\Plus\>_{B'}|^2}
\\
&\le 2 \sqrt{\sum_{m'=0,1} {}_{A'}\<\chi_{m'}|\chi_{m'}\>_{A'}}~\sqrt{\sum_{m=0,1} {}_{B'}\<\psi_m|\psi_m\>_{B'}}
\\
&\le \frac{2}{\sqrt{\left( 1- \frac{1}{2^{|B|/2}} \right)\left( 1- \frac{1}{2^{|A|/2}} \right)}}
\\
&\le 4~,
\fe
where we used the norm constraints \eqref{norm1} and \eqref{norm2} in the third inequality, and assumed that $A$ and $B$ are large enough in the last inequality. Next, we have
\ie
&\sum_{m=0,1} |{}_{A'}\<\mmm|\chi_+\>_{A'}|~|{}_{B'}\<\psi_m|\Plus\>_{B'}|
\\
&\le \sqrt{\sum_{m=0,1} |{}_{A'}\<\mmm|\chi_+\>_{A'}|^2}~\sqrt{\sum_{m=0,1} |{}_{B'}\<\psi_m|\Plus\>_{B'}|^2}
\\
&\le \sqrt{{}_{A'}\<\chi_+|\chi_+\>_{A'}}~\sqrt{\sum_{m=0,1} {}_{B'}\<\psi_m|\psi_m\>_{B'}}
\\
&\le \frac{1}{\sqrt{\left( 1- \frac{2}{2^{|B|/2}} \right)\left( 1- \frac{1}{2^{|A|/2}} \right)}}
\\
&\le 2~.
\fe
Similarly, we have
\ie
\sum_{m=0,1} |{}_{A'}\<\Plus|\chi_m\>_{A'}|~|{}_{B'}\<\psi_+|\mmm\>_{B'}| \le 2~.
\fe
And finally, we have
\ie
&\sum_{m=0,1} |{}_{A'}\<\mmm|\chi_+\>_{A'}|~|{}_{B'}\<\psi_+|\mmm\>_{B'}|
\\
&\le \sum_{m,m'=0,1} |{}_{A'}\<\mmm|\chi_+\>_{A'}|~|{}_{B'}\<\psi_+|\mmm'\>_{B'}|
\\
&\le \sqrt{2 \sum_{m=0,1} |{}_{A'}\<\mmm|\chi_+\>_{A'}|^2}~\sqrt{2 \sum_{m'=0,1} |{}_{B'}\<\psi_+|\mmm'\>_{B'}|^2}
\\
&\le 2 \sqrt{{}_{A'}\<\chi_+|\chi_+\>_{A'}}~\sqrt{{}_{B'}\<\psi_+|\psi_+\>_{B'}}
\\
&\le \frac{2}{\sqrt{\left( 1- \frac{2}{2^{|B|/2}} \right)\left( 1- \frac{2}{2^{|A|/2}} \right)}}
\\
&\le 4~.
\fe
Combining these inequalities, we have
\ie
|\<\psi|\chi\>| &\le \frac{8}{2^{(|A|+|B|)/2}} + \frac{4}{2^{|A\cap B|/2}}~.
\fe
Since this inequality holds for all $|\psi\>$ and $|\chi\>$ satisfying the constraints in \eqref{1d-delta-state-def}, we get the upper bound \eqref{1d-delta-upper-bound} on $\delta(A,B)$ for all sufficiently large $A$ and $B$.

\subsection{3+1d}\label{sec:3d-delta}
We now prove the upper bound \eqref{3d-upperbound-delta-maintext} on $\delta(A,B)$ for the Hamiltonian $H''$ \eqref{3d-deformedH''} in 3+1d.

Let $\Lambda$ be a finite connected region in the lattice containing at least one cube and with smooth boundary.\footnote{Here, $\Lambda$ is said to have a \emph{smooth boundary} if there are no dangling links or plaquettes on the boundary, i.e., every link is contained in a plaquette within $\Lambda$ and every plaquette is contained in a cube within $\Lambda$. This terminology is reminiscent of the smooth boundary of the 2+1d toric code.} We use $\partial \Lambda$ to denote the boundary of $\Lambda$, which is along the plaquettes of the lattice. Let $\mathcal H_\Lambda := \bigotimes_{\ell \in \Lambda} \mathcal H_\ell$ be the tensor product Hilbert space on $\Lambda$.

Consider the restriction of the Hamiltonian $H''$ in \eqref{3d-deformedH''} to $\Lambda$,
\ie
h_\Lambda := \sum_{\ell,p\in \Lambda:\ell\perp p} P_\ell Q_p~,
\fe
Since $\Lambda$ contains at least one cube and has a smooth boundary, using an argument similar to the one in Appendix \ref{app:deformed-nogausslaw}, we can show that $h_\Lambda$ has ground states,\footnote{Note that these states are well-defined even when $\Lambda$ does not have a smooth boundary, but they are ground states of $h_\Lambda$ only when $\Lambda$ has a smooth boundary.}
\ie
&|\Plus\>_\Lambda := |{+}\cdots{+}\>_\Lambda~,\qquad |S\>_\Lambda := \prod_{s\in S} G_{s|\Lambda} |0\cdots0\>_\Lambda~,
\fe
where $S$ is any subset of sites in $\Lambda$ and we defined the restricted Gauss law operator
\ie
G_{s|\Lambda} := \prod_{\ell \in \Lambda: \ell \ni s} X_\ell~,
\fe
Note that $|S\>_\Lambda = |\Lambda\smallsetminus S\>_\Lambda$ because of the relation $\prod_{s\in \Lambda} G_{s|\Lambda} = 1$. So there are $1+2^{V_\text{s}(\Lambda)-1}$ ground states, where $V_\text{s}(\Lambda)$ is the number of sites in $\Lambda$. While the states $|S\>_\Lambda$ are orthogonal to each other, except for complements, they are not orthogonal to $|\Plus\>_\Lambda$:
\ie
{}_\Lambda\<\Plus|S\>_\Lambda = {}_\Lambda\<{+}\cdots{+}|0\cdots0\>_\Lambda = \frac{1}{2^{V_\text{l}(\Lambda)/2}} > 0~,
\fe
where $V_\text{l}(\Lambda)$ is the number of links in $\Lambda$. We use $\mathcal H^0_\Lambda$ to denote the subspace spanned by $|\Plus\>_\Lambda$ and $|S\>_\Lambda$, and $ \Pi_\Lambda$ to denote the orthogonal projection onto $\mathcal H^0_\Lambda$, i.e., $\im( \Pi_\Lambda) = \mathcal H^0_\Lambda$.

Consider two overlapping rectangular boxes $A$ and $B$ containing at least one cube each and with smooth boundaries. Note that $A\cup B$, $A':=A\smallsetminus B$, and $B':=B\smallsetminus A$ need not be rectangular boxes, but $A\cap B$ is a rectangular box. On the other hand, $A\cup B$ and $A\cap B$ have smooth boundaries, but $A'$ and $B'$ do not. We use the following convention: the links in $\partial A \cap B'$ belong to $A$ but not $B'$, whereas the sites in $\partial A \cap B'$ belong to both $A$ and $B'$, and similarly for $A'\cap \partial B$. With this convention, the Hilbert space and the Gauss law operator factorize nicely:
\ie\label{factorization}
\mathcal H_{A\cup B} = \mathcal H_{A'} \otimes \mathcal H_{A\cap B} \otimes \mathcal H_{B'}~,\qquad G_{s|A\cup B} = G_{s|A'} G_{s|A\cap B} G_{s|B'}~.
\fe
Moreover, for any subsets $S$ and $T$ of sites in $A$ and $B$, respectively, we have
\ie\label{innerproduct}
{}_A\<\Plus|\Plus\>_B &= |\Plus\>_{B'}~{}_{A'}\<\Plus|~,
\\
{}_A\<S|\Plus\>_B &= {}_A\<0\cdots 0| \prod_{s\in S} G_{s|A} |{+}\cdots{+}\>_B
\\
&= {}_{A'}\<0\cdots 0|\prod_{s\in S\cap A'} G_{s|A'} \left({}_{A\cap B} \<0\cdots0|\prod_{s\in S\cap B} G_{s|A\cap B} |{+}\cdots{+}\>_{A\cap B} \right) |{+}\cdots{+}\>_{B'}
\\
&= \frac1{2^{V_\text{l}(A\cap B)/2}} |\Plus\>_{B'}~{}_{A'}\<S\cap A'|~,
\\
{}_A\<\Plus|T\>_B &= \frac1{2^{V_\text{l}(A\cap B)/2}} |T\cap B'\>_{B'}~{}_{A'}\<\Plus|~,
\\
{}_A\<S| T\>_B &= {}_A\<0\cdots 0| \prod_{s\in S} G_{s|A} \prod_{s'\in T} G_{s'|B} |0\cdots0\>_B
\\
&= {}_{A'}\<0\cdots 0| \prod_{s\in S\cap A'} G_{s|A'}
\\
&\quad \times \left({}_{A\cap B} \<0\cdots0|\prod_{s\in S\cap B} G_{s|A\cap B} \prod_{s'\in T\cap A} G_{s'|A\cap B} |0\cdots0\>_{A\cap B} \right)
\\
&\quad \times \prod_{s'\in T\cap B'} G_{s'|B'} |0\cdots0\>_{B'} 
\\
&= \begin{cases}
|T\cap B'\>_{B'}~{}_{A'}\<S\cap A'|~, & \text{if } S\cap B = T \cap A \text{ or } (A\cap B) \smallsetminus (T\cap A)~,
\\
0~,&\text{otherwise}~.
\end{cases}
\fe
We write $S\sim T$ for $S\subseteq A$ and $T\subseteq B$ if they satisfy the condition in the last equality.

Note that $ \Pi_A  \Pi_{A\cup B} =  \Pi_{A\cup B}  \Pi_A =  \Pi_{A\cup B}$, and similarly for $ \Pi_B$, because of frustration-freeness. It follows that $ \Pi_A -  \Pi_{A\cup B}$ and $ \Pi_B -  \Pi_{A\cup B}$ are orthogonal projections too. Frustration-freeness also implies that $ \Pi_A \wedge  \Pi_B =  \Pi_{A\cup B}$.\footnote{In general, frustration-freeness implies only that $\mathcal H^0_{A\cup B} \subseteq \mathcal H^0_A \cap \mathcal H^0_B$. For the other direction, consider any $|\psi\> \in \mathcal H^0_A \cap \mathcal H^0_B$. Then, $ \Pi_A^\perp |\psi\>=0 \implies P_\ell Q_p |\psi\> = 0$ for all $\ell,p\in A$ such that $\ell\perp p$ due to frustration-freeness, and similarly for all $\ell,p\in B$ such that $\ell \perp p$. Therefore, $P_\ell Q_p |\psi\> = 0$ for all $\ell,p\in A\cup B$ such that $\ell \perp p$, which means $ \Pi_{A\cup B}^\perp|\psi\> = 0$. There is one subtlety here: it is possible that there are $\ell \in A'$ and $p\in B'$ such that $\ell \perp p$, which means the constraint $P_\ell Q_p |\psi\> = 0$ is not obviously implied by $|\psi\> \in \mathcal H^0_A \cap \mathcal H^0_B$. However, it is still possible to show that $|\psi\> \in \mathcal H^0_{A\cup B}$ using an argument similar to the one in Section \ref{sec:3d-proofofgsd}. Therefore, $\mathcal H^0_{A\cup B} = \mathcal H^0_A \cap \mathcal H^0_B$, or equivalently, $ \Pi_A \wedge  \Pi_B =  \Pi_{A\cup B}$.}

Consider the quantity
\ie\label{3d-delta-def}
\delta(A,B) := \Vert  \Pi_A  \Pi_B -  \Pi_A \wedge  \Pi_B \Vert = \Vert  \Pi_A  \Pi_B -  \Pi_{A\cup B} \Vert = \Vert ( \Pi_A - \Pi_{A\cup B})( \Pi_B -  \Pi_{A\cup B}) \Vert~,
\fe
Since the norm of product of orthogonal projections is always at most $1$, we have $0\le \delta(A,B) \le 1$. Below, we will show that
\ie\label{3d-delta-upper-bound}
\delta(A,B) \le 4\cdot \frac{2^{V_\text{s}(A\cap B)/2}}{2^{V_\text{l}(A\cap B)/2}} + 3 \cdot \frac{2^{(V_\text{s}(A)+V_\text{s}(B))/2}}{2^{(V_\text{l}(A)+V_\text{l}(B))/2}}~,
\fe
for all sufficiently large $A$ and $B$. In order to prove this bound, we consider an equivalent characterization of $\delta(A,B)$ in terms of states instead of projections:
\ie\label{3d-delta-state-def}
\delta(A,B) &= \sup\{ |\<\psi|\chi\>|: |\psi\>,|\chi\> \in \mathcal H_{A\cup B}\,,\, \<\psi|\psi\> \le 1\,,\,\<\chi|\chi\> \le 1\,,
\\
&\qquad \qquad \qquad \quad( \Pi_A - \Pi_{A\cup B})|\psi\> =|\psi\>\,,\, ( \Pi_B - \Pi_{A\cup B})|\chi\> =|\chi\>\}~.
\fe
Note that the condition $( \Pi_A - \Pi_{A\cup B})|\psi\> =|\psi\>$ means that $|\psi\> \in (\mathcal H^0_A \otimes \mathcal H_{B'}) \cap (\mathcal H^0_{A\cup B})^\perp$, i.e., $|\psi\>$ is a ground state in $A$ and it is orthogonal to the ground states in $A\cup B$,\footnote{Indeed, $ \Pi_A |\psi\> =  \Pi_A ( \Pi_A - \Pi_{A\cup B})|\psi\> = ( \Pi_A - \Pi_{A\cup B})|\psi\> = |\psi\>$, whereas $ \Pi_{A\cup B}|\psi\> =  \Pi_{A\cup B}( \Pi_A - \Pi_{A\cup B})|\psi\> = ( \Pi_{A\cup B} - \Pi_{A\cup B})|\psi\> = 0$.} and similarly for $|\chi\>$. So we write
\ie
&|\psi\> = |\Plus\>_A |\psi_+\>_{B'} + \frac12 \sum_{S\subseteq A} |S\>_A |\psi_S\>_{B'}~,
\\
&|\chi\> = |\chi_+\>_{A'} |\Plus\>_B + \frac12 \sum_{T\subseteq B} |\chi_T\>_{A'} |S\>_B~,
\fe
for some states $|\psi_{+,S}\>_{B'} \in \mathcal H_{B'}$ and $|\chi_{+,T}\>_{A'} \in \mathcal H_{A'}$, so that $|\psi\>$ and $|\chi\>$ are ground states in $A$ and $B$, respectively. Without loss of generality, we can assume that $|\psi_S\>_{B'} = |\psi_{A\smallsetminus S}\>_{B'}$ for any $S\subseteq A$ because $|S\>_A = |A\smallsetminus S\>_A$, and similarly for $|\chi_T\>_{A'}$.

We now restrict them to be orthogonal to the ground states in $A\cup B$. First, we have 
\ie\label{ortho1}
0 = {}_{A\cup B}\<\Plus|\psi\> = {}_{B'}\<\Plus|\psi_+\>_{B'} + \frac{1}{2\cdot 2^{V_\text{l}(A)/2}} \sum_{S\subseteq A}~{}_{B'}\<\Plus|\psi_S\>_{B'}~.
\fe
Next, for any subset $R$ of sites in $A\cup B$, we have 
\ie\label{ortho2}
0 &= {}_{A\cup B}\<R|\psi\>
\\
&= {}_{A\cup B}\<0\cdots0| \prod_{s\in R} G_{s|A\cup B} |{+}\cdots{+}\>_A |\psi_+\>_{B'}
\\
&\quad + \frac12 \sum_{T\subseteq A}~{}_{A\cup B}\<0\cdots0| \prod_{s\in R} G_{s|A\cup B} \prod_{s'\in S} G_{s'|A} |0\cdots0\>_A |\psi_S\>_{B'}
\\
&= \frac{1}{2^{V_\text{l}(A)/2}}~{}_{B'}\<0\cdots0| \prod_{s\in R\cap B'} G_{s|B'} |\psi_+\>_{B'} + {}_{B'}\<0\cdots0| \prod_{s\in R\cap B'} G_{s|B'} |\psi_{R\cap A}\>_{B'}
\\
&= \frac{1}{2^{V_\text{l}(A)/2}}~{}_{B'}\<R\cap B'|\psi_+\>_{B'} + {}_{B'}\<R\cap B'|\psi_{R\cap A}\>_{B'}~,
\fe
where, in the third equality, we used the assumption $|\psi_{A\smallsetminus S}\>_{B'} = |\psi_{S\cap A}\>_{B'}$. Similarly, we have
\ie\label{ortho3}
&{}_{A\cup B}\<\Plus|\chi\> = 0 \implies {}_{A'}\<\Plus|\chi_+\>_{A'} + \frac{1}{2\cdot 2^{V_\text{l}(B)/2}} \sum_{S\subseteq B}~{}_{A'}\<\Plus|\chi_S\>_{A'} = 0~,
\\
&{}_{A\cup B}\<R|\chi\> = 0 \implies \frac{1}{2^{V_\text{l}(B)/2}}~{}_{A'}\<R\cap A'|\chi_+\>_{A'} + {}_{A'}\<R\cap A'|\chi_{R\cap B}\>_{A'} = 0~.
\fe
The constraints on the norms give\footnote{Here, we use the inequality
\ie
{}_{B'}\<\psi_+|\psi_S\>_{B'} + {}_{B'}\<\psi_S|\psi_+\>_{B'} \ge -\frac1\varepsilon~{}_{B'}\<\psi_+|\psi_+\>_{B'} - \varepsilon~{}_{B'}\<\psi_S|\psi_S\>_{B'}~,
\fe
which holds for any $\varepsilon>0$. It follows from $\Vert |\psi_+\>_{B'} + \varepsilon |\psi_S\>_{B'} \Vert \ge 0$. We then substitute $\varepsilon = 2^{V_\text{s}(A)/2}$.}
\ie\label{norm1}
1&\ge\<\psi|\psi\>
\\
&= {}_{B'}\<\psi_+|\psi_+\>_{B'} + \frac12 \sum_{S\subseteq A}~{}_{B'}\<\psi_S|\psi_S\>_{B'} + \frac{1}{2\cdot 2^{V_\text{l}(A)/2}} \sum_{S\subseteq A} \left( {}_{B'}\<\psi_+|\psi_S\>_{B'} + {}_{B'}\<\psi_S|\psi_+\>_{B'} \right)
\\
&\ge \left( 1-\frac{2^{V_\text{s}(A)/2-1}}{2^{V_\text{l}(A)/2}} \right) {}_{B'}\<\psi_+|\psi_+\>_{B'} + \frac12 \left( 1 - \frac{2^{V_\text{s}(A)/2}}{2^{V_\text{l}(A)/2}} \right) \sum_{S\subseteq A}~{}_{B'}\<\psi_S|\psi_S\>_{B'}~,
\fe
and similarly,
\ie\label{norm2}
1 \ge \left( 1-\frac{2^{V_\text{s}(B)/2-1}}{2^{V_\text{l}(B)/2}} \right) {}_{A'}\<\chi_+|\chi_+\>_{A'} + \frac12 \left( 1 - \frac{2^{V_\text{s}(B)/2}}{2^{V_\text{l}(B)/2}} \right) \sum_{T\subseteq B}~{}_{A'}\<\chi_T|\chi_T\>_{A'}~.
\fe

Now, the overlap between $|\psi\>$ and $|\chi\>$ is
\ie
\<\psi|\chi\> &= {}_{A'}\<\Plus|\chi_+\>_{A'}~{}_{B'}\<\psi_+|\Plus\>_{B'}
\\
&\quad + \frac{1}{2\cdot 2^{V_\text{l}(A\cap B)/2}} \sum_{S\subseteq A} ~{}_{A'}\<S\cap A'|\chi_+\>_{A'}~{}_{B'}\<\psi_S|\Plus\>_{B'}
\\
&\quad + \frac{1}{2\cdot 2^{V_\text{l}(A\cap B)/2}} \sum_{T\subseteq B} ~{}_{A'}\<\Plus|\chi_T\>_{A'}~{}_{B'}\<\psi_+|T\cap B'\>_{B'}
\\
&\quad + \frac14 \sum_{S\subseteq A,T\subseteq B: S\sim T} ~{}_{A'}\<S\cap A'|\chi_T\>_{A'}~{}_{B'}\<\psi_S|T\cap B'\>_{B'}~,
\fe
where we used \eqref{innerproduct}. Using the assumption that $|\chi_T\>_B = |\chi_{B\smallsetminus T}\>_B$ for any $T\subseteq B$, the last line can be simplified to
\ie
&\sum_{S\subseteq A,T\subseteq B: S\sim T} ~{}_{A'}\<S\cap A'|\chi_T\>_{A'}~{}_{B'}\<\psi_S|T\cap B'\>_{B'}
\\
&= 2 \sum_{S\subseteq A,T\subseteq B\atop S \cap B = T\cap A} ~{}_{A'}\<S\cap A'|\chi_T\>_{A'}~{}_{B'}\<\psi_S|T\cap B'\>_{B'}
\\
&= 2 \sum_{R\subseteq A\cup B} ~{}_{A'}\<R\cap A'|\chi_{R\cap B}\>_{A'}~{}_{B'}\<\psi_{R\cap A}|R\cap B'\>_{B'}~.
\fe
Then, using the constraints \eqref{ortho1}, \eqref{ortho2}, and \eqref{ortho3}, we have
\ie
\<\psi|\chi\> &= \frac{1}{4\cdot 2^{(V_\text{l}(A)+V_\text{l}(B))/2}} \sum_{S\subseteq A,T\subseteq B}~{}_{A'}\<\Plus|\chi_T\>_{A'}~{}_{B'}\<\psi_S|\Plus\>_{B'}
\\
&\quad + \frac{1}{2\cdot 2^{V_\text{l}(A\cap B)/2}} \sum_{S\subseteq A} ~{}_{A'}\<S\cap A'|\chi_+\>_{A'}~{}_{B'}\<\psi_S|\Plus\>_{B'}
\\
&\quad + \frac{1}{2\cdot 2^{V_\text{l}(A\cap B)/2}} \sum_{T\subseteq B} ~{}_{A'}\<\Plus|\chi_T\>_{A'}~{}_{B'}\<\psi_+|T\cap B'\>_{B'}
\\
&\quad + \frac{1}{2\cdot 2^{(V_\text{l}(A)+V_\text{l}(B))/2}} \sum_{R\subseteq A\cup B} ~{}_{A'}\<R\cap A'|\chi_+\>_{A'}~{}_{B'}\<\psi_+|R\cap B'\>_{B'}~.
\fe
Taking the absolute value gives
\ie
|\<\psi|\chi\>| &\le \frac{1}{4\cdot 2^{(V_\text{l}(A)+V_\text{l}(B))/2}} \sum_{S\subseteq A,T\subseteq B} |{}_{A'}\<\Plus|\chi_T\>_{A'}|~|{}_{B'}\<\psi_S|\Plus\>_{B'}|
\\
&\quad + \frac{1}{2\cdot 2^{V_\text{l}(A\cap B)/2}} \sum_{S\subseteq A} |{}_{A'}\<S\cap A'|\chi_+\>_{A'}|~|{}_{B'}\<\psi_S|\Plus\>_{B'}|
\\
&\quad + \frac{1}{2\cdot 2^{V_\text{l}(A\cap B)/2}} \sum_{T\subseteq B} |{}_{A'}\<\Plus|\chi_T\>_{A'}|~|{}_{B'}\<\psi_+|T\cap B'\>_{B'}|
\\
&\quad + \frac{1}{2\cdot 2^{(V_\text{l}(A)+V_\text{l}(B))/2}} \sum_{R\subseteq A\cup B} |{}_{A'}\<R\cap A'|\chi_+\>_{A'}|~|{}_{B'}\<\psi_+|R\cap B'\>_{B'}|~.
\fe
Let us take care of each line separately. First, we have
\ie
&\sum_{S\subseteq A,T\subseteq B} |{}_{A'}\<\Plus|\chi_T\>_{A'}|~|{}_{B'}\<\psi_S|\Plus\>_{B'}|
\\
&\le \sqrt{2^{V_\text{s}(B)} \sum_{T\subseteq B} |{}_{A'}\<\Plus|\chi_T\>_{A'}|^2}~\sqrt{2^{V_\text{s}(A)} \sum_{S\subseteq A} |{}_{B'}\<\psi_S|\Plus\>_{B'}|^2}
\\
&\le 2^{(V_\text{s}(A)+V_\text{s}(B))/2} \sqrt{\sum_{T\subseteq B} {}_{A'}\<\chi_T|\chi_T\>_{A'}}~\sqrt{\sum_{S\subseteq A} {}_{B'}\<\psi_S|\psi_S\>_{B'}}
\\
&\le \frac{2\cdot 2^{(V_\text{s}(A)+V_\text{s}(B))/2}}{\sqrt{\left( 1 - \frac{2^{V_\text{s}(B)/2}}{2^{V_\text{l}(B)/2}} \right) \left( 1 - \frac{2^{V_\text{s}(A)/2}}{2^{V_\text{l}(A)/2}} \right)}}
\\
&\le 4\cdot 2^{(V_\text{s}(A)+V_\text{s}(B))/2}~,
\fe
where we used the norm constraints \eqref{norm1} and \eqref{norm2} in the third inequality, and assumed that $A$ and $B$ are large enough in the last inequality. Next, we have
\ie
&\sum_{S\subseteq A} |{}_{A'}\<S\cap A'|\chi_+\>_{A'}|~|{}_{B'}\<\psi_S|\Plus\>_{B'}|
\\
&\le \sqrt{\sum_{S\subseteq A} |{}_{A'}\<S\cap A'|\chi_+\>_{A'}|^2}~\sqrt{\sum_{S\subseteq A} |{}_{B'}\<\psi_S|\Plus\>_{B'}|^2}
\\
&\le \sqrt{2^{V_\text{s}(A\cap B)} \sum_{T\subseteq A'} |{}_{A'}\<T|\chi_+\>_{A'}|^2}~\sqrt{\sum_{S\subseteq A} |{}_{B'}\<\psi_S|\Plus\>_{B'}|^2}
\\
&\le 2^{V_\text{s}(A\cap B)/2} \sqrt{2~{}_{A'}\<\chi_+|\chi_+\>_{A'}}~\sqrt{\sum_{S\subseteq A} {}_{B'}\<\psi_S|\psi_S\>_{B'}}
\\
&\le \frac{2 \cdot 2^{V_\text{s}(A\cap B)/2}}{\sqrt{\left( 1-\frac{2^{V_\text{s}(B)/2-1}}{2^{V_\text{l}(B)/2}} \right) \left( 1 - \frac{2^{V_\text{s}(A)/2}}{2^{V_\text{l}(A)/2}} \right)}}
\\
&\le 4 \cdot 2^{V_\text{s}(A\cap B)/2}~.
\fe
Similarly, we have
\ie
\sum_{T\subseteq B} |{}_{A'}\<\Plus|\chi_T\>_{A'}|~|{}_{B'}\<\psi_+|T\cap B'\>_{B'}| \le 4 \cdot 2^{V_\text{s}(A\cap B)/2}~.
\fe
And finally, we have
\ie
&\sum_{R\subseteq A\cup B} |{}_{A'}\<R\cap A'|\chi_+\>_{A'}|~|{}_{B'}\<\psi_+|R\cap B'\>_{B'}|
\\
&= 2^{V_\text{s}(A\cap B)} \sum_{S\subseteq A',T\subseteq B'} |{}_{A'}\<S|\chi_+\>_{A'}|~|{}_{B'}\<\psi_+|T\>_{B'}|
\\
&\le 2^{V_\text{s}(A\cap B)} \sqrt{2^{V_\text{s}(A')} \sum_{S\subseteq A'} |{}_{A'}\<S|\chi_+\>_{A'}|^2}~\sqrt{2^{V_\text{s}(B')} \sum_{T\subseteq B'} |{}_{B'}\<\psi_+|T\>_{B'}|^2}
\\
&\le 2^{V_\text{s}(A\cap B) + (V_\text{s}(A')+V_\text{s}(B'))/2} \sqrt{2~{}_{A'}\<\chi_+|\chi_+\>_{A'}}~\sqrt{2~{}_{B'}\<\psi_+|\psi_+\>_{B'}}
\\
&\le \frac{2\cdot 2^{V_\text{s}(A\cap B) + (V_\text{s}(A')+V_\text{s}(B'))/2}}{\sqrt{\left( 1- \frac{2^{V_\text{s}(B)/2-1}}{2^{V_\text{l}(B)/2}} \right)\left( 1- \frac{2^{V_\text{s}(A)/2-1}}{2^{V_\text{l}(A)/2}} \right)}}
\\
&\le 4\cdot 2^{V_\text{s}(A\cap B) + (V_\text{s}(A')+V_\text{s}(B'))/2}
\\
&= 4\cdot 2^{(V_\text{s}(A)+V_\text{s}(B))/2}~.
\fe
Combining these inequalities, we have
\ie
|\<\psi|\chi\>| &\le \frac{4\cdot 2^{(V_\text{s}(A)+V_\text{s}(B))/2}}{4\cdot 2^{(V_\text{l}(A)+V_\text{l}(B))/2}} + \frac{4 \cdot 2^{V_\text{s}(A\cap B)/2}}{2\cdot 2^{V_\text{l}(A\cap B)/2}} + \frac{4 \cdot 2^{V_\text{s}(A\cap B)/2}}{2\cdot 2^{V_\text{l}(A\cap B)/2}}
\\
&\quad + \frac{4\cdot 2^{(V_\text{s}(A)+V_\text{s}(B))/2}}{2\cdot 2^{(V_\text{l}(A)+V_\text{l}(B))/2}}
\\
&= 4\cdot \frac{2^{V_\text{s}(A\cap B)/2}}{2^{V_\text{l}(A\cap B)/2}} + 3 \cdot \frac{2^{(V_\text{s}(A)+V_\text{s}(B))/2}}{2^{(V_\text{l}(A)+V_\text{l}(B))/2}}~.
\fe
Since this inequality holds for all $|\psi\>$ and $|\chi\>$ satisfying the constraints in \eqref{3d-delta-state-def}, we get the upper bound \eqref{3d-delta-upper-bound} on $\delta(A,B)$ for all sufficiently large $A$ and $B$.

\bibliographystyle{JHEP}
\bibliography{Deformed_3+1d}

\end{document}